\documentclass{article}
\usepackage[utf8]{inputenc}
\usepackage{amsmath}
\usepackage{amsfonts}
\usepackage{authblk}
\usepackage{graphicx,epstopdf}
\usepackage{geometry}
\usepackage{hyperref}
\usepackage{blindtext}
\usepackage{comment}
\usepackage{hyperref}
\usepackage{amsthm}
\usepackage{natbib}
\usepackage{xpatch}

\usepackage[symbol]{footmisc}

\usepackage[title]{appendix}
\usepackage[normalem]{ulem}
\usepackage[linesnumbered,ruled,vlined]{algorithm2e}
\usepackage[shortlabels]{enumitem}

\usepackage{mathtools}
\DeclareMathOperator{\Hessian}{Hess}
\hypersetup{
    colorlinks=true,
    linkcolor=red,
    filecolor=magenta,      
    urlcolor=blue,
    pdftitle={Overleaf Example},
    pdfpagemode=FullScreen,
    citecolor=black,
    }
    
\providecommand{\definitionname}{Definition}
\providecommand{\propositionname}{Proposition}
\providecommand{\theoremname}{Theorem}

\newtheorem{thm}{\protect\theoremname}
\newtheorem{defn}{\protect\definitionname}
\newtheorem{prop}{\protect\propositionname}

\title{\huge{Spike formation theory in 3D flow separation}}
\author[1]{Sreejith Santhosh}
\author[1]{Haodong Qin}
\author[2,3]{Bjoern F. Klose}
\author[2]{Gustaaf B. Jacobs}
\author[4]{Jérôme Vétel}
\author[1]{Mattia Serra \thanks{Email for correspondence: \href{mailto:mserra@ucsd.edu}{mserra@ucsd.edu}}}
\affil[1]{\small{Department of Physics, University of California, San Diego, California 92093, USA}}
\affil[2]{\small{Department of Aerospace Engineering, San Diego State University, San Diego, CA 92182, USA}}
\affil[3]{\small{Institute of Test and Simulation for Gas Turbines, German Aerospace Center (DLR), Augsburg, Germany}}
\affil[4]{\small{Department of Mechanical Engineering, Polytechnique Montréal, Montréal, QC, H3C 3A7, Canada}}
\date{\today}
\begin{document}
\maketitle
\begin{abstract}
We develop a frame-invariant theory of material spike formation during flow separation over a no-slip boundary in three-dimensional flows with arbitrary time dependence. Based on the exact evolution of the largest principal curvature on near-wall material surfaces, our theory identifies fixed and moving separation. Our approach is effective over short time intervals and admits an instantaneous limit. As a byproduct, we derive explicit formulas for the evolution of the Weingarten map and the principal curvatures of any surface advected by general three-dimensional flows. The material backbone we identify acts first as a precursor and later as the centerpiece of Lagrangian flow separation. We discover previously undetected spiking points and curves where the separation backbones connect to the boundary and provide wall-based analytical formulas for their locations. We illustrate our results on several steady and unsteady flows.
	\end{abstract}

 
\section{Introduction}
Fluid flow separation is generally regarded as fluid detachment from a no-slip boundary. It is the root cause of several complex flow phenomena, such as vortex formation, wake flow, and stall, which typically reduce engineering flow devices' performance.
For a recent survey of existing literature, we refer to \cite{sudharsan2022vorticity}, \citet{serra2018} and references therein, which include \citep{prandtl1904flussigkeitsbewegung, sears1971unsteady,sears1975boundary,liu1985simple,haller2004exact,surana2008exact,Wu2015}. \textcolor{black}{
3D flow separation is challenging to analyze and visualize, and it has been subject to numerous studies since the mid 1950s. 
Inspired by dynamical systems studies by Poincar\'e, \cite{Legendre19563, Delery2001129} and \cite{Lighthill196346} pioneered 3D flow separation research. Several methods followed after these seminal works \citep{Wu20001932,Tobak198261,Simpson1996457}. 
Years later, Haller and co-workers \citep{Surana20071290, surana2006exact,surana2008exact,Jacobs20074818,Surana2008} derived an exact theory of asymptotic 3D separation in steady flows and unsteady flows with an asymptotic mean.}

\begin{figure}
		\centering
		\includegraphics[width=\textwidth]{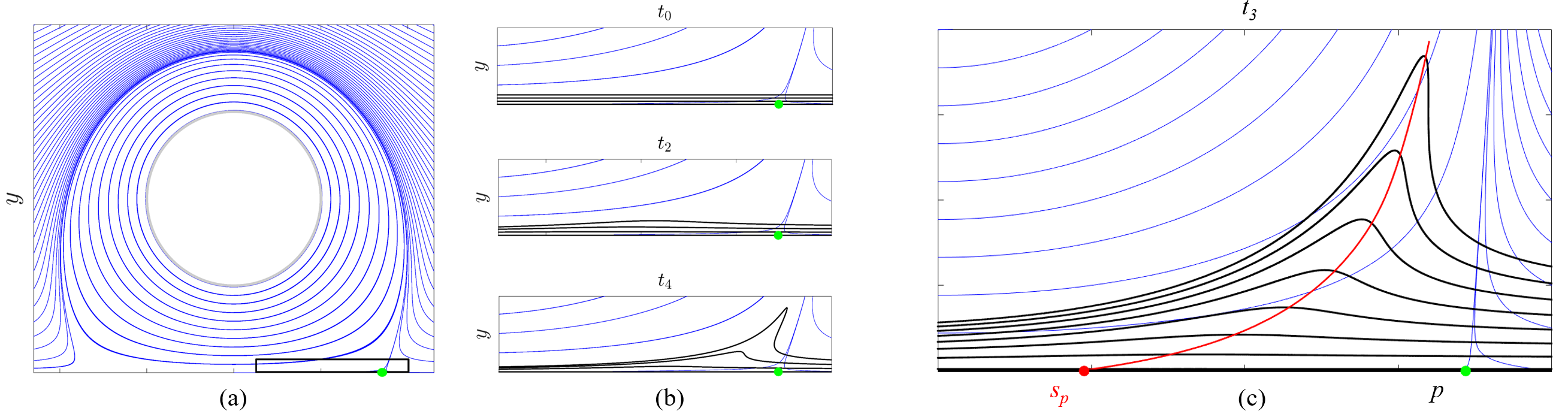}
		\caption{$\mathrm{(a)}$ Streamlines of a steady 2D flow analyzed in detail in \citet{serra2018}. The green dot represents the zero skin friction or Prandtl separation point. $\mathrm{(b)}$ Zoom of (a) in the region enclosed by the bottom right rectangle, \textcolor{black}{along with the evolution of the spike visualized through the advection of material lines shown in black. (c) Backbone of separation (red curve) as defined in \citet{serra2018}, along with the streamlines (blue), and material lines initially parallel to the wall (black). The red dot marks the spiking point, where the backbone connects to the wall. The time evolution of the material spike is in \href{https://www.dropbox.com/s/n4h93ho4abaskie/SteadyOmON_SepCurv_OnStrml_T2.mp4?dl=0}{Movie 1}.}}	
		\label{fig:2d_intro}
	\end{figure}

Existing techniques invariably focus on longer-term particle dynamics, as opposed to the appearance of separation triggered by the formation of a material spike, i.e., a sharp-shaped set of fluid particles ejected from the wall. However, long-term behavior in material deformation and transport is significantly different from short-term one, which is the most relevant for early flow separation detection and control. To illustrate the difference between short-term material spikes and longer-term material ejection, Figs. \ref{fig:2d_intro}a-b show the evolution of material lines initially close to the wall in a steady 2D flow analyzed in detail in \citet{serra2018}. While fluid particles released within the black box in (a) approach asymptotically the singular streamline (unstable manifold) emanating from the Prandtl point $p$, the birth of a material spike takes place at a different upstream location. \citet{serra2018} derived a theoretical framework to identify such a location, named the spiking point $s_p$, as well as the backbone of separation (red curve) that acts as the centerpiece of the forming spike (cf. Fig. \ref{fig:2d_intro}c and \href{https://www.dropbox.com/s/n4h93ho4abaskie/SteadyOmON_SepCurv_OnStrml_T2.mp4?dl=0}{Movie 1}) for general unsteady 2D flows. This recent technique has proven successful in identifying the onset of flow separation in highly unsteady planar flows \citep{serra2019material} including the flow over a wing profile at moderate Reynolds number \citep{Klose2019}. 

\textcolor{black}{Identifying separating structures is arguably a necessary first step in the design of flow control mechanisms \citep{YOU20081349,greenblatt2000control} that can mitigate the upwelling and break-way from walls. Common control strategies target the asymptotic separation structures either passively (\cite{schlichting00:BLT}) or actively (\cite{Cattafesta2011247}). 
 Recent efforts include optimal flow control using dynamic modes decomposition (\citep{Hemati2016,taira2017modal}) and resolvent analysis (\cite{Yeh2019572}). None of these studies, however, explicitly control spiking, and most importantly, they target Prandtl's definition of separation.} \textcolor{black}{Using the asymptotic separation criterion from \cite{haller2004exact}, \cite{kamphuis2018pulse} showed that a pulsed actuation upstream of the Haller's separation criterion reduces drag. Interestingly, \cite{bhattacharjee2020data} showed that the optimal actuator place to mitigate separation is upstream of the asymptotic separation point on an airfoil, consistent with the spiking point location \citep{Klose2020}.} A three-dimensional theory to locate and control the material spike formation universally observed in separation experiments is still missing.

 Building on \citet{serra2018}, here we propose a frame-invariant theory that identifies the origin of spike formation over a no-slip boundary in 3D flows with arbitrary time dependence. Our technique identifies the Lagrangian centerpieces--or backbone lines and surfaces (Figs. \ref{fig:intro}b,c)--of separation and is also effective over short times, which are inaccessible by previous methods. Our theory is based on explicit formulas for the Lagrangian evolution of the largest principal curvature of material surfaces. The emergence of the largest principal curvature maxima (or ridge) near the boundary locates the onset of spike formation, its dimension (1D or 2D backbones, cf Figs. \ref{fig:intro}b,c) and type. Specifically, we speak of fixed separation if the ridge emanates from the wall. Otherwise, it is a moving separation. For fixed separation, our theory discovers previously undocumented \textit{spiking points} $\mathbf{r}_{sp}$ and \textit{spiking curves} $\gamma_{sc}$, which are distinct locations where the 1D and 2D backbones connect to the wall. We provide explicit formulas for the spiking points and curves using wall-based quantities.  Remarkably, the spiking points and curves remain hidden from classic skin friction line plots even in steady flows, consistent with the 2D case (Fig. \ref{fig:2d_intro}).

This paper is organized as follows. We first develop our theoretical results in Sections \ref{sec:Setup}-\ref{sec:EulBack}. Then we give an algorithmic summary of our theory in section \ref{sec:Numerical}. In \ref{sec:Examples}, we illustrate our results on several examples, including steady and unsteady velocity fields that generate different flow separation structures over no-slip boundaries. 
\section{Set-up and notation}
	\label{sec:Setup}
	
	We consider a three-dimensional unsteady smooth velocity field $\mathbf{f}(\mathbf{x},t)$ on a three-dimensional domain $U$, whose trajectories satisfy 
	\begin{align}
	\dot{\mathbf{x}}(t) = \mathbf{f}(\mathbf{x},t), \ \ \mathbf{f}= [f_1,f_2,f_3]^\top, \ \ 
	\mathbf{x}=[x_1,x_2,x_3]^{\top} \in U \subset \mathbb{R}^3, \ \ 
	\mathbf{x}(t=0) = \mathbf{x}_0.
	\label{eq:VelODE}
	\end{align}
    We recall the customary velocity Jacobian decomposition
	\begin{equation}
	\nabla \mathbf{f}(\mathbf{x},t) = \mathbf{S}(\mathbf{x},t) + \mathbf{\Omega}(\mathbf{x},t),\quad \mathbf{S} = \frac{1}{2}(\nabla \mathbf{f} +\nabla \mathbf{f}^{\top}), 
	\label{eq:Jacobian_decomposition}
	\end{equation}
	where $\mathbf{S}$ and $\mathbf{\Omega}$ denote the rate-of-strain and the spin tensors. Trajectories $\mathbf{x}(t; t_0, \mathbf{x}_0)$ of \eqref{eq:VelODE} define the flow map $\mathbf{F}_{t_0}^{t} (\mathbf{x_0})$ and the corresponding right Cauchy–Green strain tensor $\mathbf{C}_{t_0}^{t}(\mathbf{x_0})$ that can be computed as 
	\begin{align}
	\mathbf{F}_{t_0}^{t} (\mathbf{x_0}) =   \mathbf{x_0} + \int_{t_{0}}^t \mathbf{f}(\mathbf{F}_{t_0}^{\tau}(\mathbf{x_0}),\tau)d\tau,\ \ 
	\mathbf{C}_{t_0}^{t} (\mathbf{x_0}) = [\nabla_{\mathbf{x_0}} \mathbf{F}_{t_0}^{t} (\mathbf{x_0})]^{\top} \nabla_{\mathbf{x_0}} \mathbf{F}_{t_0}^{t} (\mathbf{x_0}).
	\end{align}
	$\mathbf{F}_{t_0}^{t} (\mathbf{x}_0)$ maps an initial condition $\mathbf{x}_0$ at time $t_0$ to its position $\mathbf{x}_t$ at time $t$, and $\mathbf{C}_{t_0}^{t}(\mathbf{x_0})$ encodes Lagrangian stretching and shearing deformations of an infinitesimal material volume in the neighborhood of $\mathbf{x_0}$.
\section{Curvature evolution of a material surface}\label{sec:curvature}
	
\begin{figure}
		\centering
		\includegraphics[width=\textwidth]{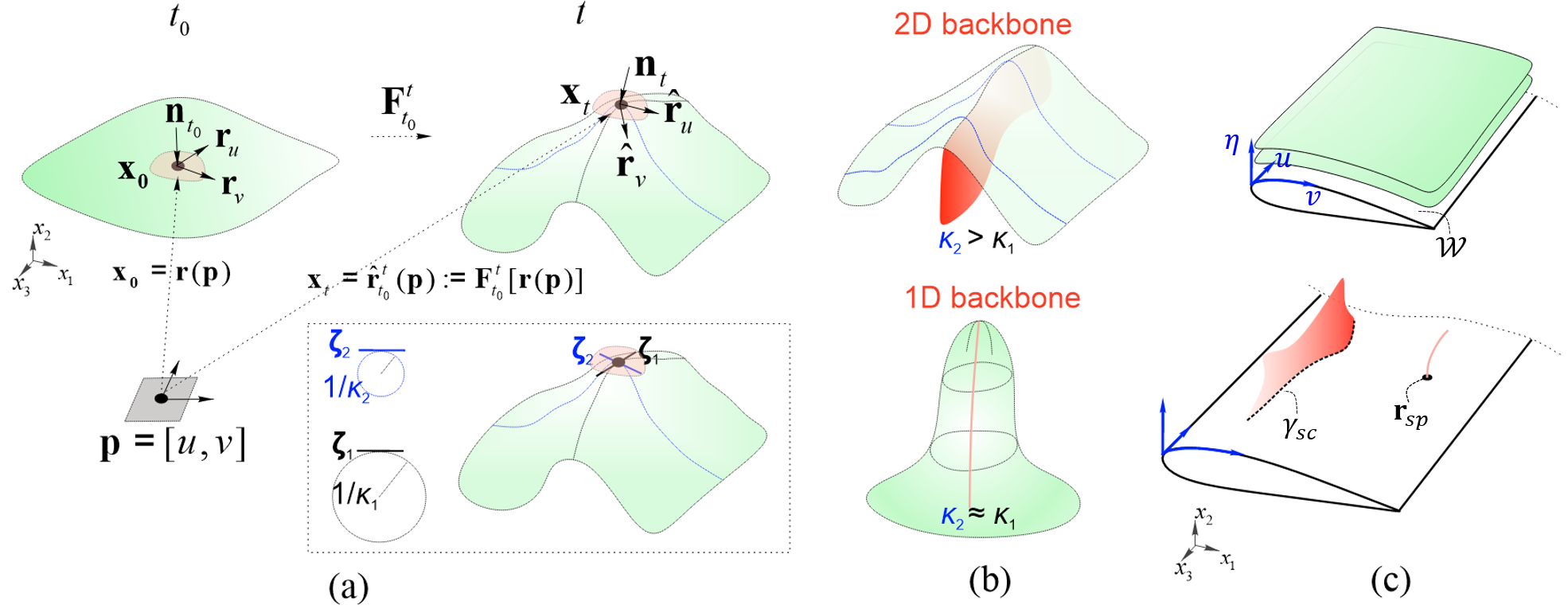}
        \caption{\textcolor{black}{(a) We denote the initial position of a two-dimensional material surface in a three-dimensional space by $\mathcal{M}(t_0)$. Points on the surface have coordinates $\mathbf{r}(\mathbf{p})$, where $\mathbf{p}$ contains the two independent parameters. At later times, points on the material surface $\mathcal{M}(t) =\mathbf{F}_{t_0}^{t} (\mathcal{M}(t_0))$ have coordinates $\hat{\mathbf{r}}_{t_0}^{t}(\mathbf{p})$. At each point $\mathbf{r}(\mathbf{p})$ of $\mathcal{M}(t)$, $\hat{\mathbf{r}}_u, \hat{\mathbf{r}}_v$ are basis vectors for the local tangent space, $\mathbf{n}_t$ is the local normal vector, $\kappa_1$ and $\kappa_2$ are the principal curvatures, and $\mathbf{\zeta}_1$ and $\mathbf{\zeta}_2$ are the principal curvature directions (inset). (b) Sketch of possible separating spikes in 3D flows from a non-slip boundary: 2D backbone where $\kappa_2>\kappa_1$ or a 1D backbone where $\kappa_2 \approx \kappa_1$. (c) We define a coordinate system with $[u,v]$ parameterizing the non-slip boundary $\mathcal{W}$ and $\eta$ the coordinate normal to the boundary. We define a spiking curve $\gamma_{sc}$ and spiking point $\mathbf{r}_{sp}$, as the intersection -- when present -- of the 2D backbone and 1D backbone to the no-slip boundary.}}
		\label{fig:intro}
	\end{figure}

	To derive explicit formulas for the curvature evolution of a two-dimensional material surface $\mathcal{M}(t)$, we define the following parametrization (Fig. \ref{fig:intro})
	\begin{equation}
\begin{split}
	    \mathcal{M}(t_0) = & \ \ \big \{\mathbf{x_0}\in U: \mathbf{x_0} =\mathbf{r}(\mathbf{p}),\  \mathbf{p}=[u,v]^{\top}\in V = [u_1,u_2]\times[v_1,v_2]\subset \mathbb{R}^2\big \},\\ \mathcal{M}(t) = & 
 \ \ \mathbf{F}_{t_0}^{t} (\mathcal{M}(t_0))=   \big \{\mathbf{x_t}\in U: \mathbf{x_t} = \hat{\mathbf{r}}_{t_0}^{t}(\mathbf{p}) =\mathbf{F}_{t_0}^{t} (\mathbf{r}(\mathbf{p})),\  \mathbf{p}\in V\big \}.\\
\end{split}
	\label{eq:M(t)}
	\end{equation}
In other words, $\mathbf{p}$ contains the two independent variables that uniquely specify an initial point of the material surface $\mathbf{r}(\mathbf{p})$. The coordinates of this point at time $t$, are given by $\mathbf{F}_{t_0}^{t} (\mathbf{r}(\mathbf{p}))$, or alternatively, in compact notation, by $ \hat{\mathbf{r}}_{t_0}^{t}(\mathbf{p})$.
 
	At each point $\hat{\mathbf{r}}_{t_0}^{t}(\mathbf{p})$ on the surface, the vectors $\partial_u \hat{\mathbf{r}}_{t_0}^{t}, \partial_v \hat{\mathbf{r}}_{t_0}^{t}$ define a basis for the local tangent space at $\hat{\mathbf{r}}_{t_0}^{t}(\mathbf{p})$. For compactness, we will denote these vectors $\mathbf{r}_u,\mathbf{r}_v$ at $t_0$ and $\hat{\mathbf{r}}_u, \hat{\mathbf{r}}_v$ at $t$. We can now compute a local basis for the local tangent and normal spaces at $\hat{\mathbf{r}}_{t_0}^{t}(\mathbf{p})$ as 
	\begin{equation}
	    	\hat{\mathbf{r}}_{u} = \nabla_{\mathbf{x_0}}\mathbf{F}_{t_0}^{t}(\mathbf{r}(\mathbf{p}))\mathbf{r}_{u},\ \hat{\mathbf{r}}_{v} = \nabla_{\mathbf{x_0}}\mathbf{F}_{t_0}^{t}(\mathbf{r}(\mathbf{p}))\mathbf{r}_{v},\ \mathbf{n}_t =\hat{\mathbf{r}}_u \times  \hat{\mathbf{r}}_v/\vert \hat{\mathbf{r}}_u \times  \hat{\mathbf{r}}_v \vert,  
	\label{eq:LocalBasis}
	\end{equation}	
	where $\times, \vert (\cdot) \vert$ denote the cross product and the vector norm, and $\mathbf{n}_t$ the unit vector normal to the surface.
	
	The Weingarten map quantifies the surface curvature in different directions, and can be computed as $\mathbf{W}_{t_0}^t(\mathbf{p}) =({}_1\mathbf{\Gamma}_{t_0}^t(\mathbf{p}))^{-1}{}_2\mathbf{\Gamma}_{t_0}^t(\mathbf{p})  $ where ${}_1\mathbf{\Gamma}$ and ${}_2\mathbf{\Gamma}$ are the first and second fundamental forms (e.g. \cite{Kuhnel2015} and Appendix \ref{sec:differential_geometry}). The eigenvalues ${}_{2}k_{t_0}^t (\mathbf{p})\geq{}_{1}k_{t_0}^t (\mathbf{p})$  of $\mathbf{W}_{t_0}^t(\mathbf{p})$ characterize the principal curvatures at $\hat{\mathbf{r}}_{t_0}^{t}(\mathbf{p})$ along the corresponding principal curvature directions $\mathbf{\zeta_1} $ and $\mathbf{\zeta_2}$, identified by the eigenvectors of $\mathbf{W}$ (Fig. \ref{fig:intro}a). As a result, the time evolution of $\mathbf{W}$ fully characterizes the curvatures of $\mathcal{M}(t)$. The Weingarten map of $\mathcal{M}(t_0)$ is $\mathbf{W}_{t_0} = \mathbf{W}_{t_0}^{t_0}(\mathbf{p})$ and can be computed as 
	
	\begin{align}
	\begin{split}
	\mathbf{W}_{t_0} = ({}_1\mathbf{\Gamma}_{t_0}(\mathbf{p}))^{-1}{}_2\mathbf{\Gamma}_{t_0}(\mathbf{p}) 
	, \ \  {}_1\mathbf{\Gamma}_{t_0}(\mathbf{p}) = \left(\begin{smallmatrix}
	\langle\mathbf{r}_u,\mathbf{r}_u\rangle&\langle\mathbf{r}_u,\mathbf{r}_v\rangle\\
	\langle\mathbf{r}_v,\mathbf{r}_u\rangle&\langle\mathbf{r}_v,\mathbf{r}_v\rangle
	\end{smallmatrix}\right),
	\ \
	{}_2\mathbf{\Gamma}_{t_0}(\mathbf{p}) =\left(\begin{smallmatrix}
	\langle\mathbf{n}_{t_0},\mathbf{r}_{uu}\rangle&\langle\mathbf{n}_{t_0},\mathbf{r}_{uv}\rangle\\
	\langle\mathbf{n}_{t_0},\mathbf{r}_{vu}\rangle&\langle\mathbf{n}_{t_0},\mathbf{r}_{vv}\rangle
	\end{smallmatrix}\right),
	\label{eq:tao_12}
	\end{split}
	\end{align} 
	with $\langle,\rangle$ denoting the dot product and $\mathbf{r}_{uv} = \partial_{uv} \mathbf{r}$. To understand how a velocity field \eqref{eq:VelODE} folds a material surface, we derive the exact map and the underlying matrix differential equation for $\mathbf{W}_{t_0}^t(\mathbf{p})$, as summarized in the following theorem.  
	\begin{thm} \label{theorem1}
		Consider a smooth material surface $\mathcal{M}(t) \subset U \subseteq \mathbb{R}^3$ parametrized at $t_0$ in the form $\mathbf{r}(\mathbf{p})$, $\mathbf{p}=[u,v]^{\top}\in [u_1,u_2]\times[v_1,v_2]=V\subset \mathbb{R}^2$, and whose tangent space is spanned by $\mathbf{r}_u(\mathbf{p})$ and $\mathbf{r}_v(\mathbf{p})$. The evolution of the Weingarten map $\mathbf{W}_{t_0}^t(\mathbf{p})$ of  $\mathcal{M}(t)$ under the action of the flow map can be computed as
		
		\begin{equation}\label{eq:Wlagrangian}
		\mathbf{W}_{t_0}^t(\mathbf{p})= \underbrace{({}_1\mathbf{\Gamma}_{t_0}^{t}(\mathbf{p}))^{-1}{}_1\mathbf{\Gamma}_{t_0}(\mathbf{p})\frac{J_{t_0}(\mathbf{p}) \textup{det}(\mathbf{\nabla}\mathbf{F}^t_{t_0}(\mathbf{r}(\mathbf{p})))}{J_{t_0}^{t}(\mathbf{p})}\mathbf{W}_{t_0}}_{\mathbf{W}_I}+\underbrace{{({}_{1}\mathbf{\Gamma}_{t_0}^t}(\mathbf{p}))^{-1}\mathbf{B}_{t_0}^{t}(\mathbf{p}),}_{\mathbf{W}_{II}}
		\end{equation}
		where
		\begin{equation}
		\mathbf{B}_{t_0}^{t}(\mathbf{p}) = \left (\begin{smallmatrix}
		\langle\mathbf{n}_t,\mathbf{\nabla}^2\mathbf{F}^t_{t_0}(\mathbf{r}(\mathbf{p}))\mathbf{r}_u\mathbf{r}_u\rangle&\langle\mathbf{n}_t,\mathbf{\nabla}^2\mathbf{F}^t_{t_0}(\mathbf{r}(\mathbf{p}))\mathbf{r}_u\mathbf{r}_v\rangle\\
		\langle\mathbf{n}_t,\mathbf{\nabla}^2\mathbf{F}^t_{t_0}(\mathbf{r}(\mathbf{p}))\mathbf{r}_v\mathbf{r}_u\rangle&\langle\mathbf{n}_t,\mathbf{\nabla}^2\mathbf{F}^t_{t_0}(\mathbf{r}(\mathbf{p}))\mathbf{r}_v\mathbf{r}_v\rangle
		\end{smallmatrix}\right),
		\end{equation}
		$\textcolor{black}{[\mathbf{\nabla}^2\mathbf{F}^t_{t_0}(\mathbf{r}(\mathbf{p})\mathbf{r}_u\mathbf{r}_v]_i = [\mathbf{\nabla}^2\mathbf{F}^t_{t_0}(\mathbf{r})\mathbf{r}_{u}]_{ij}  [\mathbf{r}_{v}]_{j}= \langle \mathbf{\nabla}[\mathbf{\nabla}\mathbf{F}^t_{t_0}(\mathbf{r})]_{ij},\mathbf{r}_{u}\rangle[\mathbf{r}_{v}]_{j}}$\footnote[2]{$[\mathbf{\nabla}^2\mathbf{F}^t_{t_0}(\mathbf{r})\mathbf{r}_{u}]_{ij}$ represents the directional derivatives of $[\mathbf{\nabla}\mathbf{F}^t_{t_0}(\mathbf{r})]_{ij}$ in the direction $\mathbf{r}_{u}$. We use the same notation in eqs. \eqref{eq:Meq}-\eqref{eq:Neq}.}, $J_{t_0}^{t}(\mathbf{p}) =\sqrt{\textup{det}({}_{1}\mathbf{\Gamma}_{t_0}^t(\mathbf{p}))}$ \textcolor{black}{ and  $J_{t_0}(\mathbf{p}) = \sqrt{\textup{det}({}_{1}\mathbf{\Gamma}_{t_0}(\mathbf{p}))}$}. The rate of change of the Weingarten map at $t_0$ is given by
		
		\begin{multline}\label{eq:WdotInit}
		\displaystyle{\dot{\mathbf{W}}_{t_0}(\mathbf{p})}= \underbrace{\displaystyle{\left[\left( \mathbf{\nabla}\cdot\mathbf{f}(\mathbf{r}(\mathbf{p}),t_0)-\frac{3\alpha_{t_0}(\mathbf{p})}{J_{t_0}^2(\mathbf{p})}\right)\mathbf{I}+\frac{2\mathbf{D}(\mathbf{p},t_0){}_1\mathbf{\Gamma}_{t_0}(\mathbf{p})}{J_{t_0}^2(\mathbf{p})}\right]\mathbf{W}_{t_0}}}_{\dot{\mathbf{W}}_{I}}\\+\underbrace{\displaystyle{({}_1\mathbf{\Gamma}_{t_0}(\mathbf{p}))^{-1}\mathbf{M}(\mathbf{p},t_0)}}_{{\dot{\mathbf{W}}_{II}}}+\underbrace{\displaystyle{({}_1\mathbf{\Gamma}_{t_0}(\mathbf{p                      }))^{-1}\mathbf{N}_{t_0}}}_{\dot{\mathbf{W}}_{III}},
		\end{multline}
		where 
		$\alpha_{t_0}(\mathbf{p}) =  \langle\mathbf{r}_{u},\mathbf{S}(\mathbf{r}(\mathbf{p}),t_0)\mathbf{r}_{u}\rangle\langle\mathbf{r}_{v},\mathbf{r}_{v}\rangle+\langle\mathbf{r}_{v},\mathbf{S}(\mathbf{r}(\mathbf{p}),t_0)\mathbf{r}_{v}\rangle\langle\mathbf{r}_{u},\mathbf{r}_{u}\rangle-2\langle\mathbf{r}_{u},\mathbf{S}(\mathbf{r}(\mathbf{p}),t_0)\mathbf{r}_{v}\rangle\langle\mathbf{r}_{u},\mathbf{r}_{v}\rangle$, $\mathbf{I}$ is the identity matrix of rank 2,

		\begin{align}
		\mathbf{D}(\mathbf{p},t_0) &=\left(\begin{smallmatrix}
		\langle\mathbf{r}_v,\mathbf{S}(\mathbf{r}(\mathbf{p}),t_0)\mathbf{r}_v\rangle&-\langle\mathbf{r}_u,\mathbf{S}(\mathbf{r}(\mathbf{p}),t_0)\mathbf{r}_v\rangle\\
		-\langle\mathbf{r}_u,\mathbf{S}(\mathbf{r}(\mathbf{p}),t_0)\mathbf{r}_v\rangle&\langle\mathbf{r}_u,\mathbf{S}(\mathbf{r}(\mathbf{p}),t_0)\mathbf{r}_u\rangle
		\end{smallmatrix}\right),\\
		\mathbf{M}(\mathbf{p},t_0) &=\left(\begin{smallmatrix}
		\langle\mathbf{n}_{t_0},\mathbf{\nabla S}(\mathbf{r}(\mathbf{p}),t_0)\mathbf{r}_u\mathbf{r}_u\rangle&\langle\mathbf{n}_{t_0},\mathbf{\nabla S}(\mathbf{r}(\mathbf{p}),t_0)\mathbf{r}_u\mathbf{r}_v\rangle\\
		\langle\mathbf{n}_{t_0},\mathbf{\nabla S}(\mathbf{r}(\mathbf{p}),t_0)\mathbf{r}_v\mathbf{r}_u\rangle&\langle\mathbf{n}_{t_0},\mathbf{\nabla S}(\mathbf{r}(\mathbf{p}),t_0)\mathbf{r}_v\mathbf{r}_v\rangle  
		\label{eq:Meq}
		\end{smallmatrix}\right),\\
		\mathbf{N}(\mathbf{p},t_0) &=
		\left(\begin{smallmatrix}
		\langle\mathbf{n}_{t_0},\mathbf{\nabla \Omega}(\mathbf{r}(\mathbf{p}),t_0)\mathbf{r}_u\mathbf{r}_u\rangle&\langle\mathbf{n}_{t_0},\mathbf{\nabla \Omega}(\mathbf{r}(\mathbf{p}),t_0)\mathbf{r}_u\mathbf{r}_v\rangle\\
		\langle\mathbf{n}_{t_0},\mathbf{\nabla \Omega}(\mathbf{r}(\mathbf{p}),t_0)\mathbf{r}_v\mathbf{r}_u\rangle&\langle\mathbf{n}_{t_0},\mathbf{\nabla \Omega}(\mathbf{r}(\mathbf{p}),t_0)\mathbf{r}_v\mathbf{r}_v\rangle
	    \label{eq:Neq}
		\end{smallmatrix}\right).\end{align}
		
		Proof. $\mathrm{See\ Appendix\ \ref{sec:differential_geometry}\ and\ Appendix\ \ref{sec:appendixB}  }$. 
	\end{thm}

 $\mathbf{W}_{I}$ in eq. \eqref{eq:Wlagrangian} describes the contribution to material folding induced by the flow if $\mathcal{M}(t_0)$ has nonzero initial curvature $\mathbf{W}_{t_0}$ while $\mathbf{W}_{II}$ folds $\mathcal{M}(t)$ regardless of $\mathbf{W}_{t_0}$. In $\mathbf{W}_{I}$,  \textcolor{black}{$J_{t_0}^t(\mathbf{p})$ relates the area element ($dA$) in $\mathcal{M}(t)$ to the area element ($dudv$) at $\mathbf{p}$ by $dA = J_{t_0}^{t}(\mathbf{p})dudv$}, $\textup{det}(\mathbf{\nabla}\mathbf{F}^t_{t_0}(\mathbf{r}(\mathbf{p}))$ accounts for volume changes, and the first fundamental form ${}_1\mathbf{\Gamma}_{t_0}(\mathbf{p})$ accounts for the shape of $\mathcal{M}(t_0)$. In $\mathbf{W}_{II}$, $\mathbf{B}_{t_0}^{t}(\mathbf{p})$ accounts for the folding of $\mathcal{M}(t)$ described by second spatial derivatives of $\mathbf{F}^t_{t_0}$. In the short-time limit, \eqref{eq:WdotInit} quantifies the rate of change of $\mathbf{W}_{t_0}^t(\mathbf{p})$ at $t_0$, and elucidates which flow features contribute to the folding rate of $\mathcal{M}_{t_0}$. $\dot{\mathbf{W}}_{I}$) encodes the compressibility of $\mathbf{f}$ ($\mathbf{\nabla}\cdot \mathbf{f}$), the stretching rate along $\mathcal{M}_{t_0}$ ($\alpha_{t_0}(\mathbf{p})$ and $\mathbf{D}(\mathbf{p},t_0)$) and the metric properties (${}_1\Gamma_{t_0}(\mathbf{p})$), weighted by its current curvature $\mathbf{W}_{t_0}$. $\dot{\mathbf{W}}_{II}$) accounts for spatial variations of the stretching rates on $\mathcal{M}_{t_0}$ encoded in $\nabla\mathbf{S}$; and $\dot{\mathbf{W}}_{III}$) accounts for spatial variations of rigid-body rotation rates on $\mathcal{M}_{t_0}$ encoded in $\mathbf{\nabla \Omega}$. Equations \eqref{eq:Wlagrangian} and \eqref{eq:WdotInit} have the same functional form as their two-dimensional analogs in \cite{serra2018} describing the folding of material curves, with the tensor $\mathbf{W}_{t_0}^t$ replacing the scalar curvature $\kappa_{t_0}^t$.
 
	Theorem \ref{theorem1} shows that the Lagrangian folding and the Eulerian folding rate of a material surface are caused by stretching- and rotation-based quantities. In Appendix {\ref{App:InvarianceProof}}, we show that $\mathbf{W}_{t_0}^t(\mathbf{p})$ and $\dot{\mathbf{W}}_{t_0}^t(\mathbf{p})$ are invariant with respect to changes in the parametrization $\mathbf{r}(\mathbf{p})$ and time-dependent rotations and translations of the coordinate frame. Remarkably, although the spin tensor $\mathbf{\Omega}$ is not objective, its spatial variations contributing to folding  (cf. eq.\eqref{eq:Neq}) is  objective, similar to the 2D case \citep{serra2018}. We summarize these results as follows.
\begin{prop}
	\label{sec:Prop1}
Denote all Euclidean coordinate changes by 
\begin{equation}
\tilde{\mathbf{x}}=\mathbf{Q}(t)\mathbf{x}+\mathbf{b}(t),
\label{eq:CoordchangeObj}
\end{equation}
where $\mathbf{Q}(t)\in SO(3)$ and $\mathbf{b}(t)\in\mathbb{R}^{3}$ are smooth functions of time. $\mathbf{W}_{t_0}^t(\mathbf{p})$ and $\dot{\mathbf{W}}_{t_0}^t(\mathbf{p})$ are independent of the parametrization $\mathbf{r}(\mathbf{p})$ (eq. \eqref{eq:M(t)}) and invariant under the coordinate changes in eq. \eqref{eq:CoordchangeObj}. Invariance here means $\tilde{\mathbf{W}}_{t_0}^t(\mathbf{p})=\mathbf{W}_{t_0}^t(\mathbf{p})$ and $\tilde{\dot{\mathbf{W}}}_{t}(\mathbf{p})=\dot{\mathbf{W}}_{t}(\mathbf{p})$, where $\tilde{(\cdot)}$ denotes quantities expressed as a function of the $\tilde{\mathbf{x}}-$coordinate, and $(\cdot)$ the same quantity expressed in terms of $\mathbf{x}-$coordinate.\\ 
\\
	\noindent Proof. $\mathrm{See\ Appendix\ {\ref{App:InvarianceProof}}.}$ \hfill\ensuremath{\square}
\end{prop}

We note that the invariance of material folding in Proposition \ref{sec:Prop1} is stronger than Objectivity \citep{TruesdellNoll2004}, which is required in the continuum mechanics assessment of material response and the definitions of Lagrangian and Eulerian Coherent Structures \citep{haller2015lagrangian,SerraHaller2015}.

\section{The Lagrangian backbone of flow separation}\label{sec:LagrBack}
	
The onset of fluid flow separation is characterized by a distinctly folded material spike that will later separate from the boundary surface, similar to the spike formation in 2D flows (Fig. \ref{fig:2d_intro}). In 3D, however, the Lagrangian backbone of separation -- i.e. the centerpiece of the material spike -- can be one-dimensional (codimension 2) or two-dimensional (codimension 1). A one-dimensional backbone marks an approximately symmetric spike. In contrast, a two-dimensional backbone marks a ridge-like spike where folding perpendicular to the ridge is higher than the one along the ridge (cf. Fig. \ref{fig:intro}b). 

Equipped with the exact expressions from Section \ref{sec:curvature}, we proceed with the definition and identification of the Lagrangian backbone of separation in 3D. We first define the Lagrangian change and the Eulerian rate of change of the Weingarten map as

  \begin{align}
	\overline{\mathbf{W}}_{t_0}^{t}(\mathbf{p}) = \mathbf{W}_{t_0}^{t}(\mathbf{p})-\mathbf{W}_{t_0}(\mathbf{p}),\quad \dot{\overline{\mathbf{W}}}_{t_0}(\mathbf{p}) = \dot{\mathbf{W}}_{t_0}(\mathbf{p}),
	\label{eq:Wchange}
	\end{align}
which quantify the finite-time folding and instantaneous folding rates induced by the flow on $\mathcal{M}(t)$. We denote the eigenvalues of $\overline{\mathbf{W}}_{t_0}^{t}(\mathbf{p})$ by  ${}_{1}\overline{\kappa}_{t_{0}}^t(\mathbf{p})\leq {}_{2}\overline{\kappa}_{t_{0}}^t(\mathbf{p})$ and the associated eigenvectors by $\overline{\mathbf{\zeta}}_1, \overline{\mathbf{\zeta}}_2$. ${}_{2}\overline{\kappa}_{t_{0}}^t(\mathbf{p})$ quantifies the highest curvature change, i.e. the folding induced by $\mathbf{F}_{t_0}^t$, at $\mathbf{r}(\mathbf{p})$. By selecting normal vectors pointing towards the non-slip boundary, positive eigenvalues of $\overline{\mathbf{W}}_{t_0}^{t}(\mathbf{p})$ mark upwelling-type deformations. 

As aggregate curvature measures described by $\overline{\mathbf{W}}_{t_0}^{t}(\mathbf{p})$, we denote the Gaussian curvature change by $\overline{K}_{t_0}^t(\mathbf{p})= \det[\overline{\mathbf{W}}_{t_0}^{t}(\mathbf{p})]={}_{1}\overline{\kappa}_{t_{0}}^t(\mathbf{p}) {}_{2}\overline{\kappa}_{t_{0}}^t(\mathbf{p})$ and the mean curvature change by $\overline{H}_{t_0}^t(\mathbf{p})= \frac{\text{Trace}[\overline{\mathbf{W}}_{t_0}^{t}(\mathbf{p})]}{2}= \frac{{}_{1}\overline{\kappa}_{t_{0}}^t(\mathbf{p})+ {}_{2}\overline{\kappa}_{t_{0}}^t(\mathbf{p})}{2}$. For compactness, we may denote the principal curvatures changes also by $\overline{\kappa}_{1},\overline{\kappa}_{2}$, the Gaussian curvature change by $\overline{K}$ and the mean curvature change by $\overline{H}$. We note that the Gaussian curvature $\overline{K}$ is not a good metric to characterize the Lagrangian spike formation because in the case of flat (or approximately flat) 2D separation ridges (e.g. Fig. \ref{fig:6plot_twisted_ridge}), $\overline{K} \approx 0$ on the ridge. Similarly, the mean curvature change $\overline{H}$ is not a good metric for separation as in the case of hyperbolic-type upwelling deformations, i.e. when $\overline{K}<0$, $\overline{H}$ can vanish on points along the separation backbone (e.g. Fig. \ref{fig:6plot_curved_ridge}).
	
We observe that for either one- and two-dimensional separation backbones, high values of $\overline {\kappa}_2$ mark the material spike location. For the two-dimensional separation backbone, $\overline{\kappa}_2$ is maximum along the principle direction $\overline{\mathbf{\zeta}}_2$ (Fig. \ref{fig:intro}a-b). In the one-dimensional separation backbones, however, $\overline{\kappa}_2 \cong \overline{\kappa}_1$ and $\overline{\mathbf{\zeta}}_1, \overline{\mathbf{\zeta}}_2$ are not defined (Fig. \ref{fig:intro}a-b). In this symmetric case, maxima of $\overline{\kappa}_2$ mark the separation backbone. To express this coherence principle mathematically, we consider a general curved no-slip boundary (Fig. \ref{fig:intro}c) and a set--mathematically, a foliation--of wall-parallel material surfaces at $t_0$ parametrized by $\mathbf{r}_{\eta}(\mathbf{p}),\  \eta\in [0,\eta_{1}],\  \eta_1\in \mathbb{R}^+$, where the boundary is defined as
	\begin{align}
	\mathcal{W} := \{ \mathbf{r}_{\eta}(\mathbf{p})\subset\mathbb{R}^3, \mathbf{p}\in V, \eta = 0 \}.
	\end{align}

	We denote the largest principle curvature change along each layer ($\eta = const.$) as ${}_{2}\overline{\kappa}_{\eta}$, the Weingarten map change as $\overline{\mathbf{W}}_{\eta}$ and the corresponding Gauss and mean curvature changes by $ \overline{K}_{\eta},\overline{H}_{\eta}$. Following \citep{serra2018}, we give the following mathematical definition.
 \vspace{12pt}
	\begin{defn} \label{sec:DefLagrBackBoneCod1}
		The Lagrangian backbone $\mathcal{B}(t)$ of separation is the theoretical centerpiece of the material spike over the time interval $[t_0,t_0+T]$. 
		
		a) A one-dimensional backbone $\mathcal{B}(t)$ is an evolving material line whose initial position $\mathcal{B}(t_0)$ is a set of points made by positive-valued maxima of the  ${}_{2}\overline{\kappa}_{\eta}$ field (Fig. \ref{fig:intro}b). For each $\eta = const.$ layer, $\mathcal{B}(t_0)$ is made of positive maximum points of ${}_{2}\overline{\kappa}_{\eta}$.
		
		b) A two-dimensional backbone $\mathcal{B}(t)$ is an evolving material surface whose initial position $\mathcal{B}(t_0)$ is a positive-valued, wall-transverse ridge of ${}_{2}\overline{\kappa}_{\eta}$ (Fig. \ref{fig:intro}b). For each $\eta = const.$ layer, $\mathcal{B}(t_0)$ is made of positive maxima of ${}_{2}\overline{\kappa}_{\eta}$ along the principle direction $\mathbf{\overline{\zeta}_2}$.
	\end{defn}
 \vspace{12pt}
To discern one- and two-dimensional separation backbones, we first identify the set of points $\mathbf{r}_{\eta}(\mathbf{p})$ on different ($\eta=const.$) layers where $2\sqrt{\vert \overline{K}_{\eta}(\mathbf{p})\vert} = \vert \overline{H}_{\eta}(\mathbf{p})\vert$. On these points, $\overline{\mathbf{W}}_{t_0}^{t}(\mathbf{p})$ does not have distinct eigenvalues. Within this set, a one-dimensional separation backbone $\mathcal{B}(t_0)$ at $t_0$ is made of positive maximum points of ${}_2\overline{\kappa}_{\eta}(\mathbf{p}):= \overline{H}_{\eta}(\mathbf{p})/2$, specified by the conditions in Proposition \ref{sec:Prop2}i left. The first condition ensures material upwelling while the second and third conditions ensure that ${}_2\overline{\kappa}_{\eta}(\mathbf{p})$ is maximum, i.e. that ${}_2\overline{\kappa}_{\eta}(\mathbf{p})$ has zero gradient and a negative definite Hessian. By contrast, two-dimensional separation backbones $\mathcal{B}(t_0)$ at $t_0$ are made of points $\mathbf{r}_{\eta}(\mathbf{p})$ on different ($\eta=const.$) layers where $2\sqrt{\vert \overline{K}_{\eta}(\mathbf{p})\vert} \neq \vert \overline{H}_{\eta}(\mathbf{p})\vert$. Within this set, $\mathcal{B}(t_0)$ is made of positive maxima of ${}_{2}\overline{\kappa}_{\eta}$ along the principle direction $\mathbf{\overline{\zeta}_2}$, specified by the conditions in Proposition \ref{sec:Prop2}i right. 
	
Similar to the 2D case \citep{serra2018}, in 3D, the points (curves) where the Lagrangian one- (two-) dimensional separation backbones connect to the wall are of particular interest for understanding whether the separation is on-wall or off-wall and for potential flow control strategies. We name these on-wall points \textit{Lagrangian spiking points $\mathbf{r}_{sp}$} and \textit{Lagrangian spiking curves $\mathbf{\gamma}_{sc}$} (Fig. \ref{fig:intro}c). They can be identified as the intersection of $\mathcal{B}(t_0)$ with the wall $\mathcal{W}$,
	\begin{equation}
	\mathbf{r}_{sp} := \text{one-dimensional} \mathcal{B}(t_0) \cap 	\mathcal{W},\ \ \gamma_{sc} := \text{two-dimensional} \mathcal{B}(t_0) \cap 	\mathcal{W}.
	\label{eq:SpikPointDef}	    
	\end{equation}

	\begin{table}
		\centering
		\begin{tabular}{c|c}
			 \multicolumn{2}{c}{For compressible flows  ($\mathbf{\nabla\cdot}\mathbf{f} \neq 0$), define $ \overline{\mathbf{W}}_{\delta\eta}(\mathbf{p}) :=  \partial_{\eta}\overline{\mathbf{W}}_\eta(\mathbf{p})\vert_{\eta=0}$.}\\
 			 \multicolumn{2}{c}{For incompressible flows  ($\mathbf{\nabla\cdot}\mathbf{f} = 0$), define $ \overline{\mathbf{W}}_{\delta\eta}(\mathbf{p}) :=  \partial_{\eta\eta}\overline{\mathbf{W}}_\eta(\mathbf{p})\vert_{\eta=0}$.}\\
			\hline 
			Lagrangian spiking point $\mathbf{r}_{sp} = \mathbf{r}_{\eta=0}(\mathbf{p}_{sp})$  & Lagrangian spiking curve $\mathbf{\gamma}_{sc}=\{\mathbf{r}_{\eta=0}(\mathbf{p}_{sc})\}$ \\
			$2\sqrt{\vert \overline{K}_{\delta\eta}(\mathbf{p})\vert} = \vert \overline{H}_{\delta\eta}(\mathbf{p})\vert$
		    & $2\sqrt{\vert \overline{K}_{\delta\eta}(\mathbf{p})\vert} \neq \vert \overline{H}_{\delta\eta}(\mathbf{p})\vert$\\
			\hline 
			${}_2\overline{\kappa}_{\delta\eta}(\mathbf{p}):= \overline{H}_{\delta\eta}(\mathbf{p})/2,$ &
			${}_2\overline{\kappa}_{\delta\eta}(\mathbf{p}):= \text{max eigenvalue} [\overline{\mathbf{W}}_{\delta\eta}(\mathbf{p})],$\\
			
			$
			\begin{cases}
			{}_2\overline{\kappa}_{\delta\eta}(\mathbf{p}_{sp}) >0,\\
			\mathbf{\nabla_p} \ {}_2\overline{\kappa}_{\delta\eta}(\mathbf{p}_{sp}) = \mathbf{0},\\
			\Hessian{[{}_2\overline{\kappa}_{\delta\eta}(\mathbf{p}_{sp})}] \prec 0.\\
			\end{cases}
			$
			&
			$
			\begin{cases}
	     	{}_2\overline{\kappa}_{\delta\eta}(\mathbf{p}_{sc}) >0,\\
			\langle \mathbf{\nabla_p} \ {}_2 \overline{\kappa}_{\delta\eta}(\mathbf{p}_{sc}), \mathbf{\overline{\zeta}_2}\rangle  = 0,\\
			\langle \mathbf{\overline{\zeta}_2}, \Hessian{[{}_2\overline{\kappa}_{\delta\eta}(\mathbf{p}_{sc})}] \mathbf{\overline{\zeta}_2}\rangle < 0.\\
			\end{cases}
			$
		\end{tabular}
		\caption{Exact criteria for the Lagrangian spiking points $\mathbf{r}_{sp}$ and curves $\mathbf{\gamma}_{sc}$ on a no-slip boundary over the time interval $[t_0,t_0+T]$ for compressible and incompressible flows. All quantities $\overline{(\cdot)}$ describe eigenvalues, eigenvectors, trace and determinant of $\overline{\mathbf{W}}_{\delta\eta}(\mathbf{p})$, consistent with our earlier notation.}
		\label{tab:sepPointLagra}
	\end{table}
	\vspace{12pt}
    \begin{table}
		\centering
		\begin{tabular}{c|c |c}
		
			 \multicolumn{3}{c}{$\mathbf{\nabla\cdot}\mathbf{f} \neq 0.$ $\quad  \overline{\mathbf{W}}_{\delta\eta}(\mathbf{p}) = \partial_{\eta}\overline{\mathbf{W}}_\eta(\mathbf{p})\vert_{\eta=0}$} \\
			\hline 
			Steady & Time-periodic: $\mathbf{f}(\mathbf{x},t+T_p) = \mathbf{f}(\mathbf{x},t)$ & Temporally aperiodic\\
			
            & $T = n T_p,\ \ n\in\mathbb{N}^+$& \\ 
            \\
            
            $ \left(\begin{smallmatrix}
		\partial_{uu\eta}f_3&\partial_{vu\eta}f_3\\
		\partial_{uv\eta}f_3&\partial_{vv\eta}f_3
		
		\end{smallmatrix}\right)$
		&
		$ \left(\begin{smallmatrix}
		\int^{t_0+T_p}_{t_0}\partial_{uu\eta}f_3dt&\int^{t_0+T_p}_{t_0}\partial_{vu\eta}f_3dt\\
		\int^{t_0+T_p}_{t_0}\partial_{uv\eta}f_3dt&\int^{t_0+T_p}_{t_0}\partial_{vv\eta}f_3dt
		
		\end{smallmatrix}\right) $
		&
		$ \left(\begin{smallmatrix}
		\int^{t_0+T}_{t_0}\partial_{uu\eta}f_3dt&\int^{t_0+T}_{t_0}\partial_{vu\eta}f_3dt\\
		\int^{t_0+T}_{t_0}\partial_{uv\eta}f_3dt&\int^{t_0+T}_{t_0}\partial_{vv\eta}f_3dt
		
		\end{smallmatrix}\right)$
		\end{tabular}
		\begin{tabular}{c|c |c}
\hline 
			 \multicolumn{3}{c}{$\mathbf{\nabla\cdot}\mathbf{f} = 0.$ $\quad  \overline{\mathbf{W}}_{\delta\eta}(\mathbf{p}) = \partial_{\eta\eta}\overline{\mathbf{W}}_\eta(\mathbf{p})\vert_{\eta=0}$} \\
			\hline 
			Steady & Time-periodic: $\mathbf{f}(\mathbf{x},t+T_p) = \mathbf{f}(\mathbf{x},t)$ & Temporally aperiodic\\
			
            & $T = n T_p,\ \ n\in\mathbb{N}^+$& \\ 
            \\
            
            $ \left(\begin{smallmatrix}
		\partial_{uu\eta,\eta}f_{3}&\partial_{vu\eta\eta}f_{3}\\
		\partial_{uv\eta\eta}f_{3}&\partial_{vv\eta\eta}f_{3}
		\end{smallmatrix}\right)$
		&
		$ \left(\begin{smallmatrix}
		\int^{t_0+T_p}_{t_0}\partial_{uu\eta\eta}f_3dt&\int^{t_0+T_p}_{t_0}\partial_{vu\eta\eta}f_3dt\\
		\int^{t_0+T_p}_{t_0}\partial_{uv\eta\eta}f_3dt&\int^{t_0+T_p}_{t_0}\partial_{vv\eta\eta}f_3dt
		\end{smallmatrix}\right) $
		&
		$ \left(\begin{smallmatrix}
		\int^{t_0+T}_{t_0}\partial_{uu\eta\eta}f_3dt&\int^{t_0+T}_{t_0}\partial_{vu\eta\eta}f_3dt\\
		\int^{t_0+T}_{t_0}\partial_{uv\eta\eta}f_3dt&\int^{t_0+T}_{t_0}\partial_{vv\eta\eta}f_3dt\end{smallmatrix}\right)$
		\end{tabular}
		\caption{Formulas for computing $\overline{\mathbf{W}}_{\delta\eta}(\mathbf{p})$ used in the definitions of the Lagrangian spiking points and curves (Table \ref{tab:sepPointLagra}) in terms of on-wall Eulerian quantities for steady, time-periodic and time aperiodic flows. \textcolor{black}{Here $f_3= f_3(\mathbf{r}_\eta(\mathbf{p}),t)$, and derivatives are evaluated at $\eta= 0$.}}
		\label{tab:WallWeingEulerian}
	\end{table}
We provide below an alternative method for locating $\mathbf{r}_{sp}$ and $\mathbf{\gamma}_{sc}$ in terms of the Weingarten map. Because ${}_2\overline{\kappa}_{0}(\mathbf{p}) = 0$ on the wall, $\mathbf{r}_{sp}$ and $\mathbf{\gamma}_{sc}$ are distinguished wall points and lines with maximal positive ${}_2\overline{\kappa}_{\eta}(\mathbf{p})$ in the limit of $\eta \rightarrow 0$. \textcolor{black}{To this end, we define $\overline{K}_{\delta\eta}(\mathbf{p}) = \text{det}(\overline{\mathbf{W}}_{\delta\eta}(\mathbf{p}))$\footnote[2]{We use the subscript $0<\delta \eta \ll 1$ to indicate the leading order contribution of material folding close to the wall.} and $\overline{H}_{\delta\eta}(\mathbf{p}) = \text{Tr}(\overline{\mathbf{W}}_{\delta\eta}(\mathbf{p}))$, where $\overline{\mathbf{W}}_{\delta\eta}(\mathbf{p})$ encodes the leading order curvature change close to the wall. Using $\overline{K}_{\delta\eta}$ and $\overline{H}_{\delta\eta}$, in Appendix \ref{App:LagrSpikingPoint} we derive explicit formulae for the Lagrangian spiking points and curves in the case of compressible and incompressible flows. The only difference between the two cases is that in the former $\overline{\mathbf{W}}_{\delta\eta}(\mathbf{p}):=\partial_{\eta}\overline{\mathbf{W}}_\eta(\mathbf{p})\vert_{\eta=0}$, while in the latter $ \overline{\mathbf{W}}_{\delta\eta}(\mathbf{p}) :=  \partial_{\eta\eta}\overline{\mathbf{W}}_\eta(\mathbf{p})\vert_{\eta=0}$. We summarize our results for the identification of $\mathbf{r}_{sp}$ and $\mathbf{\gamma}_{sc}$  in terms of Lagrangian quantities in Table \ref{tab:sepPointLagra}.}	
	
In Table \ref{tab:WallWeingEulerian}, we provide exact formulas for computing  $\overline{\mathbf{W}}_{\delta\eta}(\mathbf{p})$ used in the definitions of the Lagrangian spiking points and curves (Table \ref{tab:sepPointLagra}) in terms of on-wall Eulerian quantities for steady, time-periodic and time aperiodic flows. 
The formulas in Tables \ref{tab:sepPointLagra}-\ref{tab:WallWeingEulerian} highlight three important facts. First, in the case of steady flows, spiking points and curves are fixed, independent of $T$, and can be computed from derivatives of the velocity field on the wall. Second, in the case of $T_p$-periodic flows, with $T$ equal to any arbitrary multiple of $T_p$, spiking points and curves are fixed, independent of $t_0$, and can be computed by averaging derivatives of the velocity field on the wall over one period. Third, for general unsteady flows or time-periodic flows with $T\neq nT_p,\ n\in\mathbb{N}^+$, spiking points and curves move depending on $t_0$ and $T$, and can be computed by averaging derivatives of the velocity field over $[t_0,t_0+T]$. We summarize the results of this section in the following Proposition.
\vspace{18 pt}
	\begin{prop}\label{sec:Prop2}
		Over the finite-time interval $t\in [t_0, t_0+T]$:
		\begin{enumerate}[label=(\alph*)]
			\item{The initial position $\mathcal{B}(t_0)$  of the Lagrangian backbone of separation can be computed as the set of points $\mathbf{r}_{\eta}(\mathbf{p})\in U, \mathbf{p}\in V, \eta \in [0,\eta_1]$ that satisfy the following conditions. }

			\makeatletter
				\let\@float@original\@float
				\xpatchcmd{\@float}{\csname fps@#1\endcsname}{h!}{}{}
				\makeatother
					\begin{table}
		\centering
		\begin{tabular}{c|c}
			1D Lagrangian backbone of separation    & 2D Lagrangian backbone of separation\\
			$2\sqrt{\vert \overline{K}_{\eta}(\mathbf{p})\vert} = \vert \overline{H}_{\eta}(\mathbf{p})\vert$
		    & $2\sqrt{\vert \overline{K}_{\eta}(\mathbf{p})\vert} \neq \vert\overline{H}_{\eta}(\mathbf{p})\vert$\\
			\hline 
	     	${}_2\overline{\kappa}_{\eta}(\mathbf{p}):= \overline{H}_{\eta}(\mathbf{p})/2,$ &
			\textcolor{black}{${}_2\overline{\kappa}_{\eta}(\mathbf{p}):= \text{max eigenvalue} [\overline{\mathbf{W}}_{t_0}^{t}(\mathbf{p})],$}\\ 
			$
			\mathcal{B}(t_0):=
			\begin{cases}
			{}_2\overline{\kappa}_{\eta}(\mathbf{p}) >0,\ \ \ \ \ \ \ \  \ \ \ \ \  \eta \in (0,\eta_1]\\
			\mathbf{\nabla_p} \ {}_2\overline{\kappa}_{\eta}(\mathbf{p}) = \mathbf{0},\ \ \ \ \ \ \ \  \eta \in (0,\eta_1]\\
			\Hessian{[{}_2\overline{\kappa}_{\eta}(\mathbf{p})}] \prec 0 , \ \ \eta \in (0,\eta_1]\\
			(\mathbf{r}_{sp},\eta),  \ \ \ \ \ \ \ \ \ \ \ \ \ \ \ \ \ \ \ \ \ \ \eta = 0,\\
			\end{cases}
			$
			&
			$
			\mathcal{B}(t_0):=
			\begin{cases}
	     	{}_2\overline{\kappa}_{\eta}(\mathbf{p}) >0,\ \ \ \ \ \ \ \ \ \ \ \ \ \ \ \ \ \ \ \ \ \ \ \ \ \ \  \eta \in (0,\eta_1]\\
			\langle \mathbf{\nabla_p} \ {}_2 \overline{\kappa}_{\eta}(\mathbf{p}), \mathbf{\overline{\zeta}_2}\rangle  = 0,\ \ \ \ \ \ \ \ \ \ \ \ \  \eta \in (0,\eta_1]\\
			\langle \mathbf{\overline{\zeta}_2}, \Hessian{[{}_2\overline{\kappa}_{\eta}(\mathbf{p})}] \mathbf{\overline{\zeta}_2}\rangle < 0, \ \ \ \eta \in (0,\eta_1]\\
			(\mathbf{\gamma}_{sc},\eta), \ \ \ \ \ \ \ \ \ \ \ \ \ \ \ \ \ \ \ \ \ \ \ \ \ \ \ \ \ \ \ \ \ \ \ \ \eta = 0,\\
			\end{cases}
			$
		   \end{tabular} 
	\end{table}
		\makeatletter
		\let\@float\@float@original
		\makeatother
			The Lagrangian spiking points $\mathbf{r}_{sp}$ and curves $\mathbf{\gamma}_{sc}$ can be computed in terms of Lagrangian quantities using the formulae in Table \ref{tab:sepPointLagra}, in terms of wall-based averaged Eulerian quantities using the formulae in Tables \ref{tab:sepPointLagra}-\ref{tab:WallWeingEulerian}, or as the intersection of $\mathcal{B}(t_0)$ with the no-slip boundary (eq. \ref{eq:SpikPointDef}).
			\item {Later positions $\mathcal{B}(t)$ of the Lagrangian backbone of separation can be computed as 
			$ \mathcal{B}(t):= \mathbf{F}_{t_0}^t(\mathcal{B}(t_0))$.}
			\item {The Lagrangian spiking points and curves 
				\makeatletter
				\let\@float@original\@float
				\xpatchcmd{\@float}{\csname fps@#1\endcsname}{h!}{}{}
				\makeatother
				\begin{table*}
					\centering
						\begin{tabular}{c|c|c}
							$Steady\ flow$ & $Time-periodic\ flow:\ \  \mathbf{f}(\mathbf{x},t+T_p)=\mathbf{f}(\mathbf{x},t)$ & $Aperiodic\ flow$ \\
							\hline 	 
							$are\ fixed$ & $if$ $T=nT_p,\ \ n\in\mathbb{N}^+$; $are\ fixed$ &	$move$ \\
							$and\ independent\ of$ $t_0,T$ & $and\ independent\ of$ $t_0,n$ & $depending\ on$ $t_0,T$.\\
					\end{tabular}
			\end{table*}
		\makeatletter
		\let\@float\@float@original
		\makeatother}
		\end{enumerate}	
	\end{prop}
 \vspace{12pt}
\noindent By Proposition \ref{sec:Prop1}, the Lagrangian backbone of separation is invariant under coordinate ($\mathbf{x}$) transformations (cf eq. \eqref{eq:CoordchangeObj}) and changes in the parametrization ($\mathbf{r}(\mathbf{p})$) of initial conditions. Although the analytic formulas in Tables \ref{tab:sepPointLagra}-\ref{tab:WallWeingEulerian} involve higher derivatives of the velocity field, the spiking point can also be identified as the intersection of $\mathcal{B}(t_0)$ with the wall (cf. eq. \ref{eq:SpikPointDef}) with low numerical effort.

\section{The Eulerian backbone of flow separation}\label{sec:EulBack}
Over an infinitesimally short-time interval, the Eulerian backbone of flow separation acts as the centerpiece of the material spike formation. We define this Eulerian concept by taking the time derivative of the Lagrangian backbone of separation and evaluating it at $T=0$. From eq. \eqref{eq:Wchange}, the rate of change of $\overline{\mathbf{W}}_{t_0}^{t}(\mathbf{p})$ in the infinitesimally-short time limit is $\dot{\mathbf{W}}_{t}(\mathbf{p})$ (eq. \eqref{eq:WdotInit}). Denoting by ${}_{1}\dot{\kappa}_{t}(\mathbf{p})\leq {}_{2}\dot{\kappa}_{t}(\mathbf{p})$, $\dot{\mathbf{\zeta}}_1, \dot{\mathbf{\zeta}}_2$ the eigenvalues and eigenvectors of $\dot{\mathbf{W}}_{t}(\mathbf{p})$, and by $\dot{K}_{t}(\mathbf{p})= \det[\dot{\mathbf{W}}_{t}(\mathbf{p})]={}_{1}\dot{\kappa}_{t}(\mathbf{p}) {}_{2}\dot{\kappa}_{t}(\mathbf{p})$, $\dot{H}_{t}(\mathbf{p})= \frac{\text{Trace}[\dot{\mathbf{W}}_{t}(\mathbf{p})]}{2}= \frac{{}_{1}\dot{\kappa}_{t}(\mathbf{p})+ {}_{2}\dot{\kappa}_{t}(\mathbf{p})}{2}$ the Gaussian and mean curvature rates, we define the Eulerian backbone of separation as follows. We may omit the explicit time dependence notation for compactness.
\vspace{12pt}
	\begin{defn} \label{sec:DefEulBone}
		At a time instant $t$, the Eulerian backbone of separation $\mathcal{B}_E(t)$ is the theoretical centerpiece of the material spike over an infinitesimally-short time interval. 
		
		a) A one-dimensional backbone $\mathcal{B}_E(t)$ is a set of points made by positive-valued maxima of the ${}_{2}\dot{\kappa}_{\eta}$ field. For each $\eta = const.$ layer, $\mathcal{B}_E(t)$ is made of positive maximum points of ${}_{2}\dot{\kappa}_{\eta}$. 
		
		b) A two-dimensional backbone $\mathcal{B}_E(t)$ is a positive-valued, wall-transverse ridge of ${}_{2}\dot{\kappa}_{\eta}$. For each $\eta = const.$ layer, $\mathcal{B}_E(t)$ is made of positive maxima of ${}_{2}\dot{\kappa}_{\eta}$ along the principle direction $\dot{\mathbf{\zeta}}_2$.
	\end{defn}
 \vspace{12pt}
\noindent $\mathcal{B}_E(t)$ is a set of points where the instantaneous folding rate is positive and attains a local maximum along each $\eta=$const. surfaces, and can be computed as described in Proposition \ref{sec:Prop3}. 

\begin{table}[h]
		\centering
		\begin{tabular}{c|c}
			 \multicolumn{2}{c}{For compressible flows  ($\mathbf{\nabla\cdot}\mathbf{f} \neq 0$), define $ \dot{\mathbf{W}}_{\delta\eta}(\mathbf{p}) :=  \partial_{\eta}\dot{\mathbf{W}}_\eta(\mathbf{p})\vert_{\eta=0}$.}\\
 			 \multicolumn{2}{c}{For incompressible flows  ($\mathbf{\nabla\cdot}\mathbf{f} = 0$), define $ \dot{\mathbf{W}}_{\delta\eta}(\mathbf{p}) :=  \partial_{\eta\eta}\dot{\mathbf{W}}_\eta(\mathbf{p})\vert_{\eta=0}$.}\\
			\hline 
			Eulerian spiking point $\mathbf{r}_{spE} = \mathbf{r}_{\eta=0}(\mathbf{p}_{spE})$  & Eulerian spiking curve $\mathbf{\gamma}_{scE}=\{\mathbf{r}_{\eta=0}(\mathbf{p}_{scE})\}$ \\
			$2\sqrt{\vert \dot{K}_{\delta\eta}(\mathbf{p})\vert} = \vert \dot{H}_{\delta\eta}(\mathbf{p})\vert$
		    & $2\sqrt{\vert \dot{K}_{\delta\eta}(\mathbf{p})\vert} \neq \vert \dot{H}_{\delta\eta}(\mathbf{p})\vert$\\
			\hline 
			${}_2\dot{\kappa}_{\delta\eta}(\mathbf{p}):= \dot{H}_{\delta\eta}(\mathbf{p})/2,$ &
			${}_2\dot{\kappa}_{\delta\eta}(\mathbf{p}):= \text{max eigenvalue} [\dot{\mathbf{W}}_{\delta\eta}(\mathbf{p})],$\\
			
			$
			\begin{cases}
			{}_2\dot{\kappa}_{\delta\eta}(\mathbf{p}_{spE}) >0,\\
			\mathbf{\nabla_p} \ {}_2\dot{\kappa}_{\delta\eta}(\mathbf{p}_{spE}) = \mathbf{0},\\
			\Hessian{[{}_2\dot{\kappa}_{\delta\eta}(\mathbf{p}_{spE})}] \prec 0.\\
			\end{cases}
			$
			&
			$
			\begin{cases}
	     	{}_2\dot{\kappa}_{\delta\eta}(\mathbf{p}_{scE}) >0,\\
			\langle \mathbf{\nabla_p} \ {}_2 \dot{\kappa}_{\delta\eta}(\mathbf{p}_{scE}), \dot{\mathbf{\zeta}_2}\rangle  = 0,\\
			\langle \dot{\mathbf{\zeta}_2}, \Hessian{[{}_2\dot{\kappa}_{\delta\eta}(\mathbf{p}_{scE})}] \dot{\mathbf{\zeta}_2}\rangle < 0.\\
			\end{cases}
			$
		\end{tabular}
		\caption{Exact criteria determining the Eulerian spiking points $\mathbf{r}_{spE}$ and curves $\mathbf{\gamma}_{scE}$ on a no-slip boundary at a time instant $t$ for compressible and incompressible flows. All quantities $\dot{\overline{(\cdot)}}$ describe eigenvalues, eigenvectors, trace and determinant of $\dot{\overline{\mathbf{W}}}_{\delta\eta}(\mathbf{p})$, consistent with our earlier notation.}
		\label{tab:sepPointEul}
	\end{table}
 \vspace{12 pt}
	\begin{table}
		\centering
		\begin{tabular}{c|c}
	     $\mathbf{\nabla\cdot}\mathbf{f} \neq 0$  &  $\mathbf{\nabla\cdot}\mathbf{f} = 0$  \\
			\hline 
    $\partial_{\eta}\dot{\mathbf{W}}_{t_0}^{t}(\mathbf{p})\vert_{\eta=0} = \left (\begin{smallmatrix}
    \partial_{uu\eta}f_3&\partial_{vu\eta}f_3\\
		\partial_{uv\eta}f_3&\partial_{vv\eta}f_3
    \end{smallmatrix}\right )$
			&
			$
    \partial_{\eta\eta}\dot{\mathbf{W}}_{t_0}^{t}(\mathbf{p})\vert_{\eta=0} = \left (\begin{smallmatrix}
    \partial_{uu\eta\eta}f_3&\partial_{vu\eta\eta}f_3\\
		\partial_{uv\eta\eta}f_3&\partial_{vv\eta\eta}f_3
    \end{smallmatrix}\right )
			$
		\end{tabular}
    \caption{Formulas for computing $\dot{\mathbf{W}}_{\delta\eta}(\mathbf{p})$ used in the definitions of the Eulerian spiking points and curves (Table \ref{tab:sepPointEul}) in terms of on-wall Eulerian quantities. Here $f_3= f_3(\mathbf{r}_\eta(\mathbf{p}),t)$, and derivatives are evaluated at $\eta= 0$ and $t= t_0$.}
		\label{tab:sepPointEulVelocities}
	\end{table}
Similar to the Lagrangian case, we define the \textit{Eulerian spiking point} $\mathbf{r}_{spE}$ and the \textit{Eulerian spiking curve} $\gamma_{scE}$
	\begin{equation}
	\mathbf{r}_{spE} := \text{one-dimensional}\  \mathcal{B}_E(t_0) \cap 	\mathcal{W},\ \ \gamma_{scE} := \text{two-dimensional}\  \mathcal{B}_E(t_0) \cap 	\mathcal{W},
	\label{eq:EulSpikPointDef}	    
	\end{equation}
i.e where the Eulerian backbones of separation connects to the wall. Because $\dot{\kappa}_{t}(\mathbf{p})\equiv0$ on the no-slip boundary, $\mathbf{r}_{spE},\ \gamma_{scE}$ are distinguished points on the wall with positive maximal curvature rate in the limit of $\eta\rightarrow0$.
\\
For a flat wall, we derive analytic expressions for $\mathbf{r}_{spE} = \mathbf{r}_{\eta=0}(\mathbf{p}_{spE})$ and $\gamma_{scE}$ (a set of  $\mathbf{r}_{scE} = \mathbf{r}_{\eta=0}(\mathbf{p}_{scE})$) in Appendix \ref{app:SepPtOnWallEul}, and summarize them in Tables \ref{tab:sepPointEul}-\ref{tab:sepPointEulVelocities}. For steady flows, comparing the formula of $\mathbf{p}_{sp}$ and $\mathbf{p}_{sc}$ (cf. Table \ref{tab:sepPointLagra}) with the one of $\mathbf{p}_{spE}$ and $\mathbf{p}_{scE}$ (cf. Table \ref{tab:sepPointEul}), we obtain that the Lagrangian and the Eulerian backbones of separation connect to the wall at the same location, i.e., $\mathbf{p}_{spE}\equiv \mathbf{p}_{sp}$ and $\gamma_{spE}\equiv \gamma_{sp}$ (see e.g., \textcolor{black}{Fig. \ref{fig:FlowPastCubeP5}}).
We summarize the results of this section in the following Proposition.
\vspace{12pt}
\begin{prop}\label{sec:Prop3}
		At a time instant $t$:
		\begin{enumerate}[(a)]
			\item {The Eulerian backbone of separation $\mathcal{B}_E(t)$ can be computed as the set of points $\mathbf{r}_{\eta}(\mathbf{p})\in U, \, \mathbf{p}\in V,\, \eta \in [0,\eta_1]$ that satisfy the following conditions. 

			\makeatletter
				\let\@float@original\@float
				\xpatchcmd{\@float}{\csname fps@#1\endcsname}{h!}{}{}
				\makeatother
					\begin{table}
		\centering
		\begin{tabular}{c|c}
			1D Eulerian backbone of separation    & 2D Eulerian backbone of separation\\
			$2\sqrt{\vert \dot{K}_{\eta}(\mathbf{p})\vert} = \vert \dot{H}_{\eta}(\mathbf{p})\vert$
		    & $2\sqrt{\vert \dot{K}_{\eta}(\mathbf{p})\vert} \neq \vert\dot{H}_{\eta}(\mathbf{p})\vert$\\
			\hline 
	     	${}_2\dot{\kappa}_{\eta}(\mathbf{p}):= \dot{H}_{\eta}(\mathbf{p})/2,$ &
			\textcolor{black}{${}_2\dot{\kappa}_{\eta}(\mathbf{p}):= \text{max eigenvalue} [\dot{\mathbf{W}}_{t_0}^{t}(\mathbf{p})],$}\\ 
			$
			\mathcal{B}_E(t):=
			\begin{cases}
			{}_2\dot{\kappa}_{\eta}(\mathbf{p}) >0,\ \ \ \ \ \ \ \  \ \ \ \ \  \eta \in (0,\eta_1]\\
			\mathbf{\nabla_p} \ {}_2\dot{\kappa}_{\eta}(\mathbf{p}) = \mathbf{0},\ \ \ \ \ \ \ \  \eta \in (0,\eta_1]\\
			\Hessian{[{}_2\dot{\kappa}_{\eta}(\mathbf{p})}] \prec 0 , \ \ \eta \in (0,\eta_1]\\
			(\mathbf{r}_{sp},\eta),  \ \ \ \ \ \ \ \ \ \ \ \ \ \ \ \ \ \ \ \ \ \ \eta = 0,\\
			\end{cases}
			$
			&
			$
			\mathcal{B}_E(t):=
			\begin{cases}
	     	{}_2\dot{\kappa}_{\eta}(\mathbf{p}) >0,\ \ \ \ \ \ \ \ \ \ \ \ \ \ \ \ \ \ \ \ \ \ \ \ \ \ \  \eta \in (0,\eta_1]\\
			\langle \mathbf{\nabla_p} \ {}_2 \dot{\kappa}_{\eta}(\mathbf{p}), \dot{\mathbf{\zeta}_2}\rangle  = 0,\ \ \ \ \ \ \ \ \ \ \ \ \  \eta \in (0,\eta_1]\\
			\langle \dot{\mathbf{\zeta}_2}, \Hessian{[{}_2\dot{\kappa}_{\eta}(\mathbf{p})}] \dot{\mathbf{\zeta}_2}\rangle < 0, \ \ \ \eta \in (0,\eta_1]\\
			(\mathbf{\gamma}_{sc},\eta), \ \ \ \ \ \ \ \ \ \ \ \ \ \ \ \ \ \ \ \ \ \ \ \ \ \ \ \ \ \ \ \ \ \ \ \ \eta = 0.\\
			\end{cases}
			$
		   \end{tabular} 
	\end{table}
		\makeatletter
		\let\@float\@float@original
		\makeatother
			The Eulerian spiking points $\mathbf{r}_{spE}$ and curves $\mathbf{\gamma}_{scE}$ can be computed using the formulas in Tables \ref{tab:sepPointEul}-\ref{tab:sepPointEulVelocities}, or as the intersection of $\mathcal{B}_E(t)$ with the no-slip boundary (eq. \ref{eq:EulSpikPointDef}).}
	
			\item {The Eulerian spiking point and curve coincides with the Lagrangian spiking point and curve in steady flows.}
		\end{enumerate}	
	\end{prop}
\vspace{12pt}
By Proposition \ref{sec:Prop1}, the Eulerian backbone of separation is objective. Following the same argument of section \ref{sec:LagrBack}, \textcolor{black}{although the analytic formulae in Table \ref{tab:sepPointEulVelocities} involve higher derivatives of the velocity field, the spiking point can also be identified with low numerical effort directly from eq. \eqref{eq:EulSpikPointDef}, as the intersection of $\mathcal{B}_E(t)$ with the wall.}

\section{Numerical schemes}\label{sec:Numerical}
We summarise the numerical steps necessary to locate Lagrangian and Eulerian separation backbones in a general three-dimensional flow. 

	\begin{algorithm}[H] 
	\caption{Compute the Lagrangian backbone $\mathcal{B}(t)$ of separation (Proposition {\ref{sec:Prop2}})}
	\textbf{Inputs:} A three-dimensional velocity field $\mathbf{f}(\mathbf{x},t)$ around a no-slip boundary over a finite-time interval $[t_0,t_0+T]$. Geometry of the no-slip boundary parametrized by $\mathbf{r}_{\eta=0}(\mathbf{p})$, $\mathbf{p}\ \in V \subset \mathbb{R}^2$.
 
			\textbf{Procedure:} Initialize a set of material surfaces parallel to the wall, parametrized in the form $\mathbf{r}_{\eta}(\mathbf{p})$, where $ \mathbf{p} \in V$, $\eta$ $ \in[0,\eta_1]$, $\eta_1>0$. Advect the material surface under the velocity field $\mathbf{f}(\mathbf{x},t)$ for the time interval $[t_0,t_0+T]$.\\
			 
			Compute the change in Weingarten map $\overline{\mathbf{W}}_{t_0}^{t_0+T}$ of the material surfaces using \eqref{eq:Wlagrangian} or using the first ${}_1\Gamma_{\eta}(\mathbf{p})$ and second ${}_2\Gamma_{\eta}(\mathbf{p})$ fundamental form as in Appendix \ref{sec:differential_geometry}. \\
			
			Compute the eigenvalues and eigenvectors of $\overline{\mathbf{W}}_{t_0}^{t_0+T}$ and identify the initial position $\mathcal{B}(t_0)$ using Proposition \ref{sec:Prop2}.\\
			
			Compute later positions of the Lagrangian backbone of separation $\mathcal{B}(t)$ by advecting its initial position $\mathcal{B}(t_0)$ under the flow map $F_{t_0}^t,\ t\in[t_0,t_0+T]$. \\
		\textbf{Output:} Lagrangian backbone of separation $\mathcal{B}(t),\ \ t \in [t_0,t_0+T].$
		\label{algorithm1}
	\end{algorithm}

	\begin{algorithm}[H] 
		\caption{Compute the Eulerian backbone $\mathcal{B}_E(t)$ of separation (Proposition {\ref{sec:Prop3}})}
	\textbf{Inputs:} A three-dimensional velocity field $\mathbf{f}(\mathbf{x},t)$ around a no-slip boundary at time $t$. Geometry of the no-slip boundary parametrized by $\mathbf{r}_{\eta=0}(\mathbf{p})$, $\mathbf{p}\ \in V \subset \mathbb{R}^2$.
		
			\textbf{Procedure:} Initialize a set of material surfaces parallel to the wall, parametrized in the form $\mathbf{r}_{\eta}(\mathbf{p})$, where $ \mathbf{p} \in V$, $\eta$ $ \in[0,\eta_1]$, $\eta_1>0$.\\
			 
			Compute the rate of change in Weingarten map $\dot{\mathbf{W}}(t)$ using \eqref{eq:WdotInit}. \\
			
			Compute the Eulerian backbone of separation $\mathcal{B}_E(t)$ using Proposition \ref{sec:Prop3}.\\
			
		\textbf{Output:} Eulerian backbone of separation $\mathcal{B}_E(t).$
		\label{algorithm2}
	\end{algorithm}

 \section{Examples}\label{sec:Examples} 
 We illustrate our results by applying Algorithms 1-2 to 3D analytical and simulated flow fields. In Sections \ref{sec:TwistedSeparationRidge}-\ref{sec:OnWallOffWall}, we introduce our results on simple, synthetic, analytical flows, demonstrating how our method captures simultaneous 1D and 2D backbones of separation. We also show that other  metrics, such as the Gaussian curvature $\Bar{{K}_{t_0}^{t}}$ and mean curvature $\Bar{H}_{t_0}^{t}$ changes are suboptimal to capture flow separation. In Section \ref{sec:OnWallOffWall}, we capture separating structures that transition from a purely on-wall separation to on-wall and off-wall separation for longer time intervals. In Sections \ref{sec:FlowPastMountedCube}-\ref{sec:MovingLSB}, we apply our results to steady and unsteady velocity fields that solve the Navier-Stokes equations and are computed by direct numerical simulation (DNS). 
 	
\subsection{Two-dimensional separation ridge curved on the wall}\label{sec:TwistedSeparationRidge}
\begin{figure}[h]
		\centering
		\includegraphics[width=\textwidth]{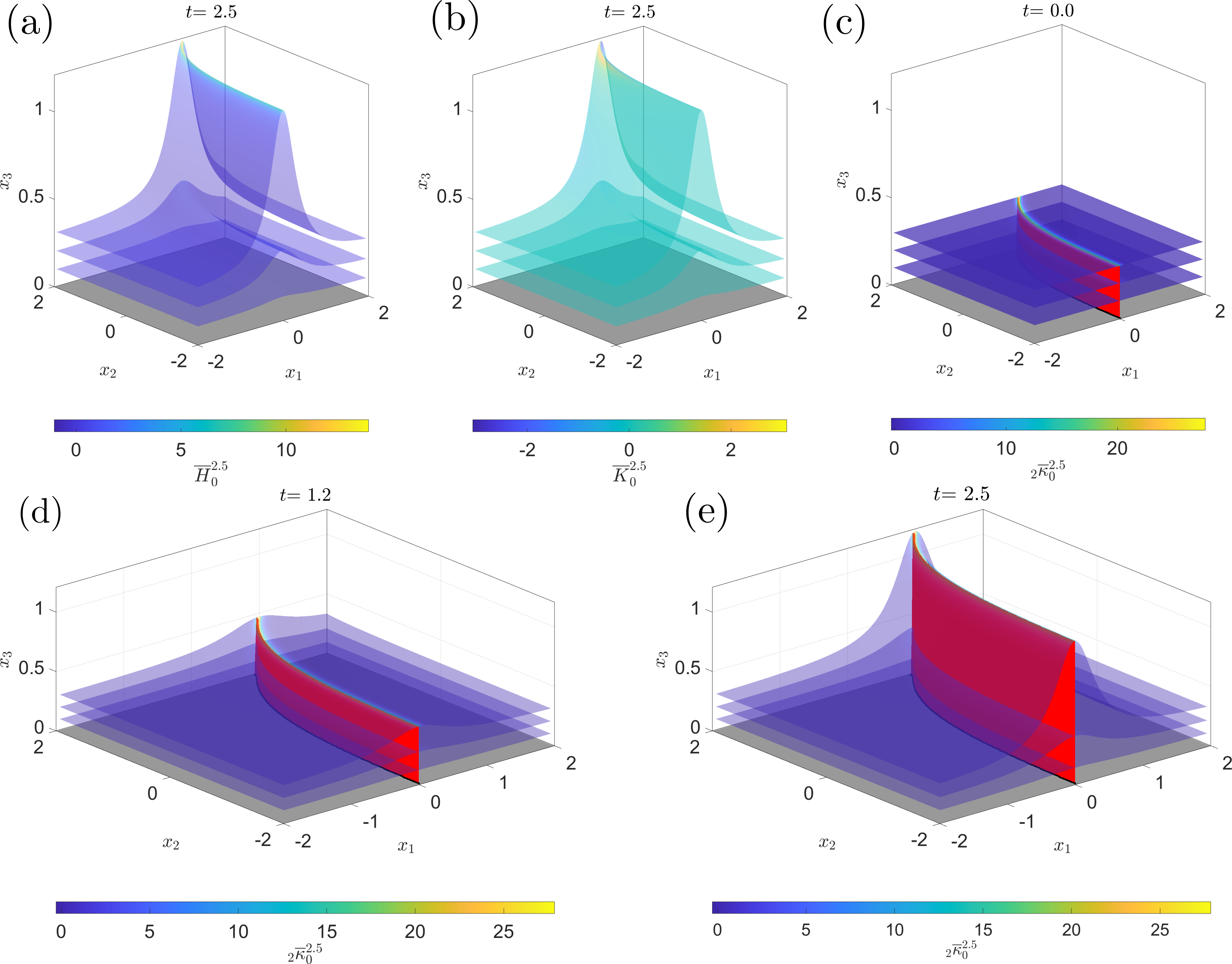}
		\caption{Separation ridge curved in $x_1-x_2$ generated by flow described in Sec. \ref{sec:TwistedSeparationRidge} over a time-interval $[0,2.5]$. (a) Mean curvature change $\overline{H}_{0}^{2.5}$ and (b) Gaussian curvature change $\overline{K}_{0}^{2.5}$ fields shaded on representative material surfaces at $t=2.5$. (c) Larger principal curvature change ${}_2\overline{\kappa}_{0}^{2.5}$ field shaded on a material surface at $t=0$. The Lagrangian backbone of separation $\mathcal{B}(0)$ is shown in red, and the black line corresponds to the Lagrangian spiking curve $\gamma_{sc}$. (d, e) The Lagrangian backbone of separation $\mathcal{B}(t)$ at later times in red, with the change in the larger principal curvature ${}_2\overline{\kappa}_{0}^{2.5}$ shaded on selected material surfaces. The time evolution of panels a-e is available in \href{https://drive.google.com/file/d/14axFtyQ_RF4smiXUDfvVLo6f1xVLey5G/view?usp=sharing}{Movie2.}}
		\label{fig:6plot_twisted_ridge}
	\end{figure}

	 We consider an analytical velocity field that generates a curved separation structure from a flat, no-slip boundary located at $x_3 = 0$. We construct the velocity field as $f_1 = 0, f_2 = 0$ and $f_3 =0.5 \frac{x_3^2}{(x_1- e^{x_2-2})^2+0.5}$, and perform a Lagrangian analysis in the time interval $[0,2.5]$.
$\overline{H}_{0}^{2.5}$ and $\overline{K}_{0}^{2.5}$ are shaded on three representative material surfaces  $\mathcal{M}(t)$ at increasing distances from the wall in Figs.\ref{fig:6plot_twisted_ridge} a-b.
Figure \ref{fig:6plot_twisted_ridge}c shows the initial Lagrangian backbone of separation $\mathcal{B}(0)$ (red) along with the largest principal curvature change field ${}_2\overline{\kappa}_{0}^{2.5}$ shaded on $\mathcal{M}(0)$. The advected backbone of separation $\mathcal{B}(t)$ is shown in Fig.\ref{fig:6plot_twisted_ridge} (d) at $t = 1.2$ and (e) at $t = 2.5$, along with ${}_2\overline{\kappa}_{0}^{2.5}$ shaded on $\mathcal{M}(t)$. The black curve on the wall in panels c-e marks the Lagrangian spiking curve, i.e., the on-wall signature of the separation backbone. The Gaussian curvature change is zero along the ridge, making $\overline{K}_{t_0}^{t_0+T}$ an unsuitable metric to characterize the separation backbone.
	\subsection{Two-dimensional separation ridge curved off-wall}\label{sec:CurvedSeparationRidge}
\begin{figure}
		\centering
		\includegraphics[width=\textwidth]{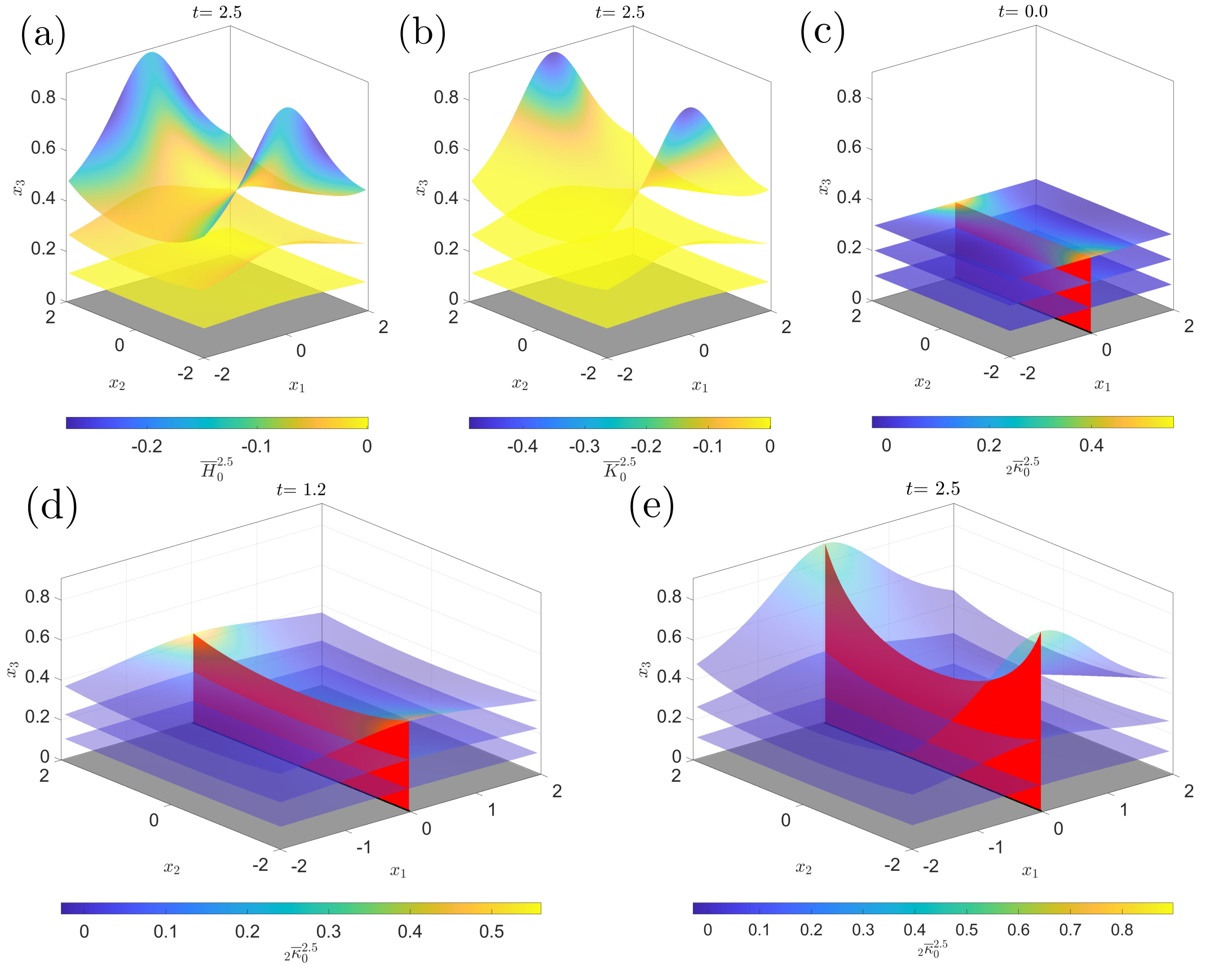}
		\caption{Separation ridge curved in the $x_2-x_3$ plane, generated by the flow described in Sec. \ref{sec:CurvedSeparationRidge} over a time-interval $[0,2.5]$. (a-d) Panels description is the same as Figure \ref{fig:6plot_twisted_ridge}. The time evolution of panels (a-d) is available in \href{https://drive.google.com/file/d/14OcYEqmd7FrOkR49toG4e7UaYRRWq6wz/view?usp=sharing}{Movie3}.}
		\label{fig:6plot_curved_ridge}
	\end{figure}

	 We consider an analytical velocity field that generates an off-wall curved separation ridge from a flat, no-slip boundary located at $x_3 = 0$. We construct a velocity field as $f_1 = 0,f_2 =0$ and $f_3= 0.5 \frac{(0.2 x_2^2+1)x_3^2}{0.2x_1^2 +1}$.
	Figure \ref{fig:6plot_curved_ridge} shows the same quantities of Figure \ref{fig:6plot_twisted_ridge} for this new velocity. $\overline{H}_{0}^{2.5}$ is zero at the saddle point of the separation spike, showing that the mean curvature change is suboptimal for identifying the backbone of separation. By contrast, ${}_2\overline{\kappa}_{0}^{2.5}$ has a maximal ridge along the centerpiece of the material spike, correctly identifying the backbone of separation (red), and the corresponding Lagrangian spiking curve $\gamma_{sc}$ (black).
	
	\subsection{Coexisting 1D and 2D backbones of separation}\label{sec:Coexisting1DAnd2D}
\begin{figure}
		\centering
		\includegraphics[width=\textwidth]{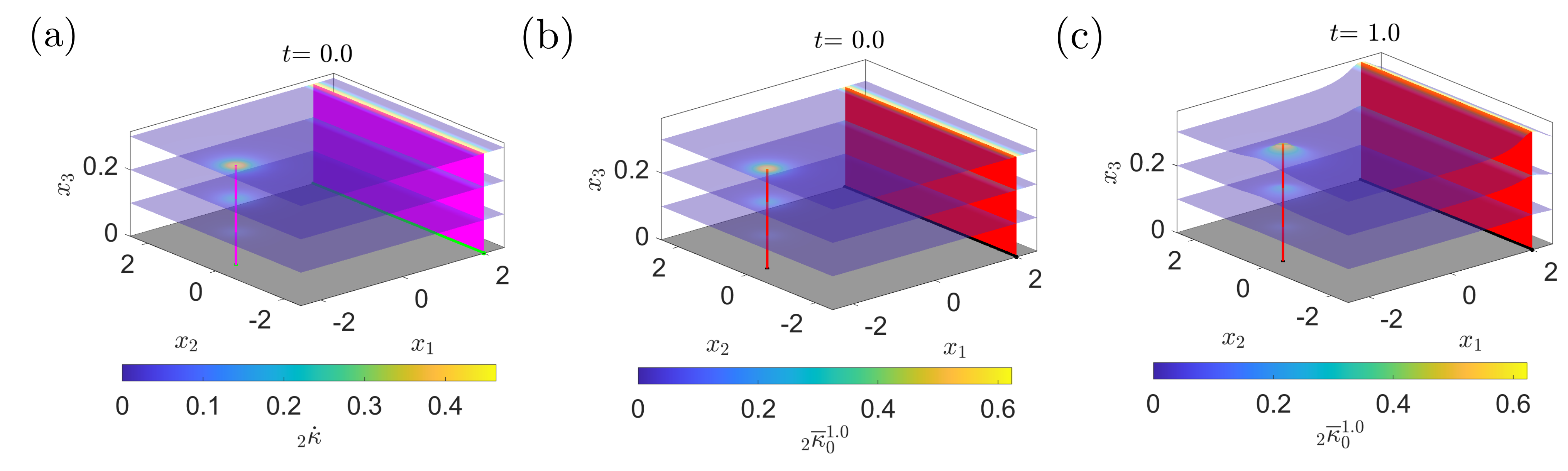}
		\caption{Coexisting 1D and 2D separation backbones generated by the flow described in Sec. \ref{sec:Coexisting1DAnd2D}. (a) Larger principal curvature rate field ${}_2\dot{\kappa}$, Eulerian backbones of separation $\mathcal{B}_E$ (magenta) and their corresponding Eulerian spiking point $\mathbf{r}_{spE}$ and curve $\gamma_{scE}$ (green).  b-c) The Lagrangian backbones of separation $\mathcal{B}(t)$(red), their corresponding Lagrangian spiking point $\mathbf{r}_{sp}$ and curve $\gamma_{sc}$ (black), along with the larger principal curvature change ${}_2\overline{\kappa}_{0}^{5}$ shaded on representative material surfaces at $t=0$ and $t = 5$. The time evolution of the above panels, along with the $\bar{H}$ and $\bar{K}$ metrics, are available in  \href{https://drive.google.com/file/d/1uoRYCWjcI0KwpPR7AFd-1BBO2_SXv1rr/view?usp=sharing}{Movie4}.}
		\label{fig:6plot_bump_ridge}
	\end{figure}

Material spike formation could lead to either 1D or 2D backbones of separation (Fig. \ref{fig:intro}). Here, we consider an analytical velocity field coexisting 1D and 2D separation backbones. The velocity field is given by $f_1= 0,f_2 = 0$ and $f_3 = 0.1 \frac{x_3^2}{(x_1+2)^2 + x_2^2 +0.2}  + 0.1 \frac{x_3^2}{(x_1-2)^2  +0.2}$. Figure \ref{fig:6plot_bump_ridge}a shows the Eulerian backbones of separation (magenta), the largest principal curvature rate ${}_2\dot{\kappa}$ shaded on representative material surfaces at different distances from the wall, and the Eulerian spiking curve $\gamma_{scE}$ and point $\gamma_{spE}$ (green), representing the on-wall footprints of the Eulerian backbones of separation. We also compute the Lagrangian backbones of separation $\mathcal{B}(t)$ over a time-interval $[0, 5]$, and show them in red along with the ${}_2\overline{\kappa}_{0}^{5}$ field at the initial and final times along their corresponding $\gamma_{sp}$ and $\gamma_{sc}$ in black (Figs. \ref{fig:6plot_bump_ridge}b-c). A closer inspection of panels a-b shows that the Eulerian and Lagrangian spiking points and curves coincide in steady flows, as predicted theoretically in Proposition \ref{sec:Prop3}. 
 
\subsection{On-wall to off-wall separation}\label{sec:OnWallOffWall}
\begin{figure}
	\centering 
	\includegraphics[height=0.24\columnwidth]{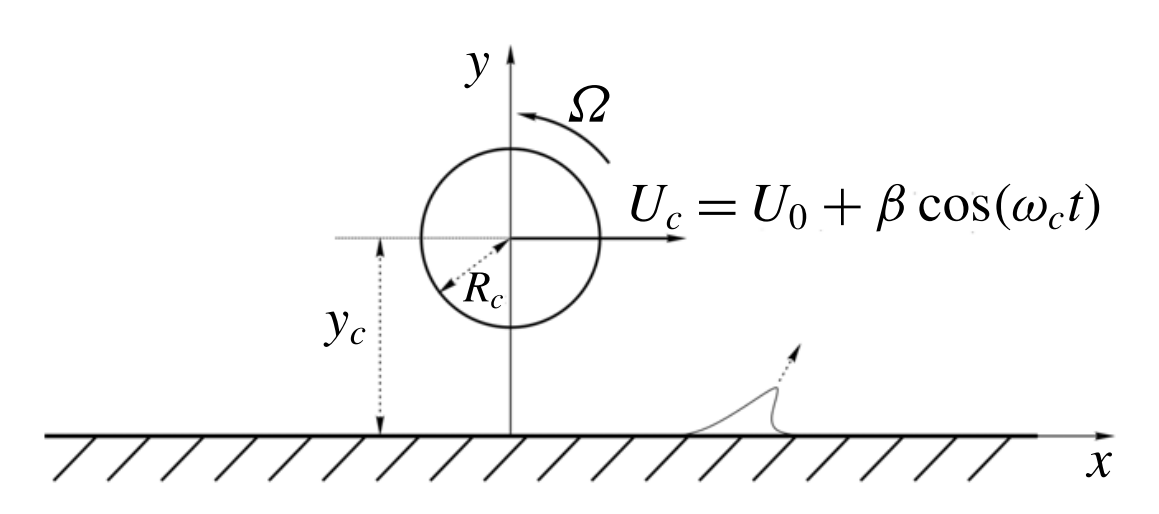}
	\caption{Setup of a flow separation induced by a rotating and translating cylinder. The analytical solution of the creeping flow generated by this setup is in Appendix \ref{App:Creepingflow}. }	
	\label{fig:CreepingFlowSetup}
    \end{figure}

\begin{figure}
		\centering
		\includegraphics[width=\textwidth]{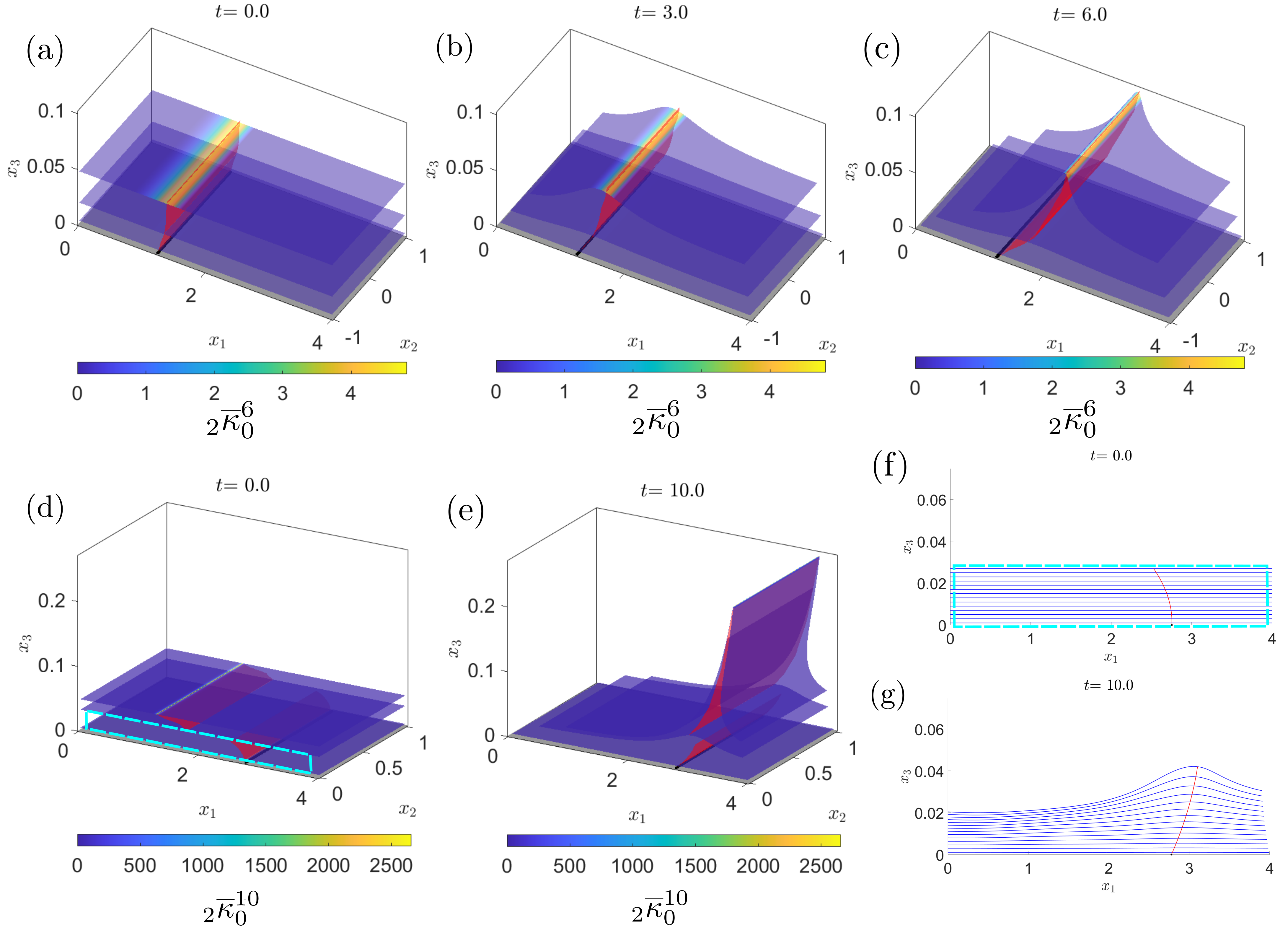}
		\caption{Separation backbone generated by the flow described in Sec. \ref{sec:OnWallOffWall}. (a-c) Change in the larger principal curvature ${}_2\overline{\kappa}_{0}^{T}$ for $T=6$ shaded on a material surface at $t=0$, $t=3$ and $t=6$. (d-e) Same as a-c for $T=10$ shaded on material surfaces at $t=0$ and $t=10$. The Lagrangian backbone of separation $\mathcal{B}(t)$ is in red, and the Lagrangian spiking curve $\gamma_{sc}$ is in black. (f-g) Evolution of the material surfaces (blue) and the separation backbone (red) intersecting the $x_1-x_3$ plane in the close-to-wall region marked by the cyan rectangle in (d). The time evolution of the above panels, along with the $\bar{H}$ and $\bar{K}$ fields, are available in \href{https://drive.google.com/file/d/14gLdWnEMa9hlS9BgNkbOHJETUZOoSs00/view?usp=sharing}{Movie5} for $T=6$ and in \href{https://drive.google.com/file/d/14dVmjf4dM_I7WBAmqz2GTkz9A7EHbfm3/view?usp=sharing}{Movie6} for $T=10$.}
		\label{fig:onwall_2plot}
	\end{figure}

We consider a general unsteady flow generated by a rotating-translating cylinder, as shown in Fig. \ref{fig:CreepingFlowSetup} with the parameters $\Omega = 3.5$, $U_0 =0.3$, $\beta=0.5$ and $\omega_c =2\pi/5$. We extend the analytical solution to this creeping 2D flow with components $[u,v]$ given in Appendix \ref{App:Creepingflow} to generate a 3D velocity field given by $f_1 = u, f_2 = 0$ and $f_3 =v$. The initial position of the Lagrangian backbone of separation $\mathcal{B}(t_0)$ computed over a time interval $T =6$ is shown in red (Fig.\ref{fig:onwall_2plot}a), along with $\gamma_{sc}$ (black) and the ${}_2\kappa_{0}^{6}$ field shaded on selected material surfaces. $\mathcal{B}(t)$ and the advected material surfaces are shown in panel b at $t=3$ and c at $t = 6$. Over this time interval, the separation is fixed or on-wall, as a single continuous backbone connects to the wall at the spiking curve $\gamma{sc}$ (black). By contrast, for a longer time scale, $ T = 10$, the Lagrangian backbone of separation $\mathcal{B}(t_0)$ becomes discontinuous (Fig. \ref{fig:onwall_2plot}d-e).
    
Because the rotating cylinder moves towards larger $x_1$ values, for longer time scales $T = 10$, the separation backbone loses its original footprint on the wall ($\gamma_{sc}$ in panels a-c) and develops two disconnected pieces. The lower backbone connecting to the wall uncovers a new spiking curve ($\gamma_{sc}$ in panels d-g), which serves as the on-wall footprint of the latest separation structure close to the wall. Figures \ref{fig:onwall_2plot}f-g, show how the lower section of the $\mathcal{B}(t)$ acts as the centerpiece of the new spike formation close to the wall. In this case, we speak about moving separation. 
The upper part, in contrast, connects to the highest value of ${}_2\overline{\kappa}$ and acts as the centerpiece of the separation spike governed by off-wall dynamics. This transition from on-wall (fixed) to off-wall (moving) separation and the ability to capture distinct separation structures over time is an automated outcome of our method -- grounded on a curvature-based theory -- which doesn't require any apriori assumptions. For a detailed discussion comparing our curvature-based theory and previous approaches to identify off-wall flow separation, see Sec. 6.2.1 of \cite{serra2018}. 

The analytical, synthetic examples above show how our approach automatically i) captures 1D and 2D separation backbones, ii) discerns on-wall (fixed) to off-wall (moving) separation, and iii) locates previously unknown on-wall signatures of fixed separation. We now apply our approach to physical flow fields.  

\subsection{Steady flow past a  cube}\label{sec:FlowPastMountedCube}
\begin{figure}
    \centering
    \includegraphics[width=.6\textwidth]{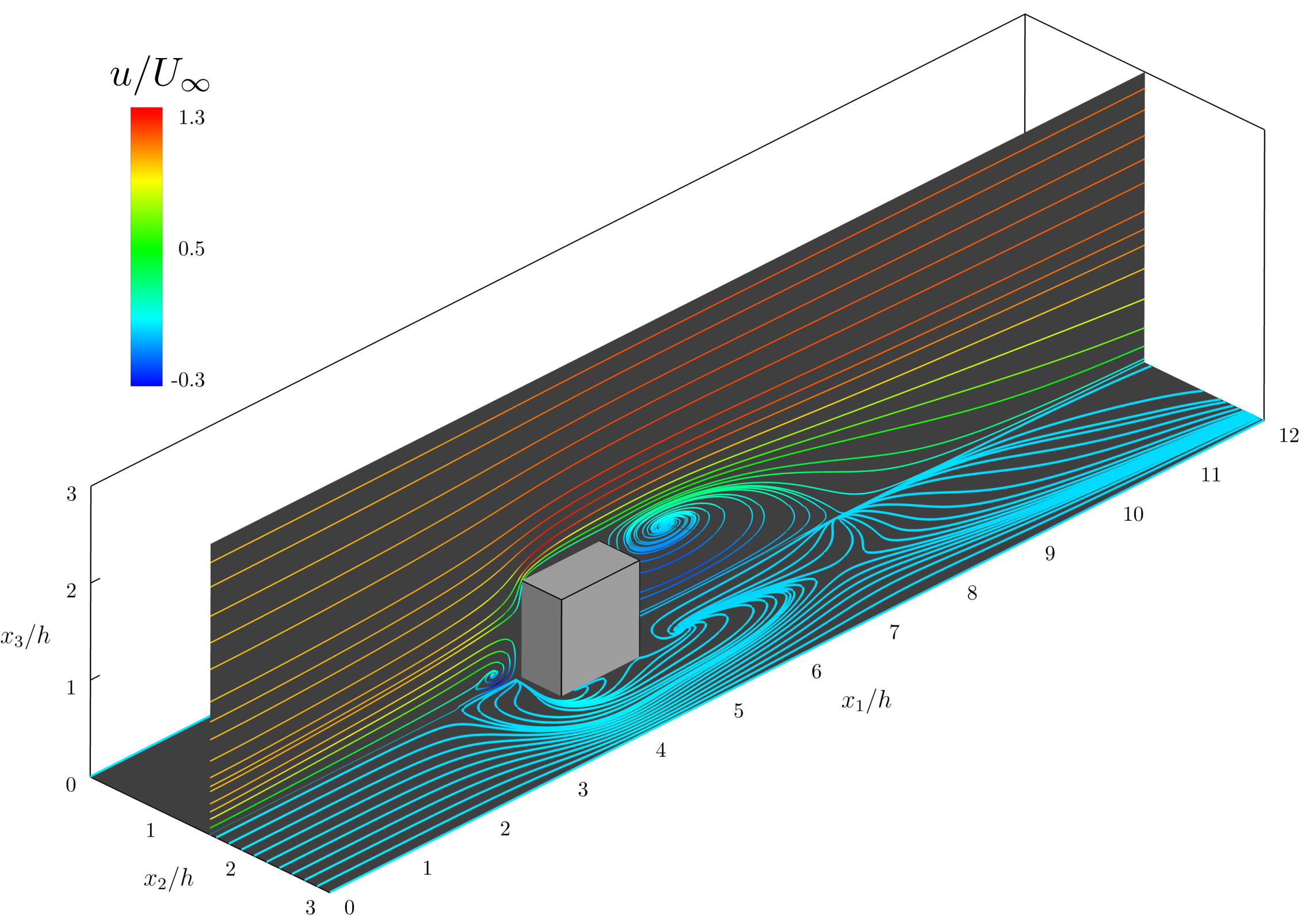}
    \caption{Steady velocity field for the flow past a mounted cube described in Appendix \ref{sec:SteadyFlowCubeDetails}. The streamlines are colored by the streamwise velocity component $u/U_{\infty}$ in the $x_2 = 1.5h$ plane and the streamlines at $x_3 = 0$ correspond to the skin friction lines. $U_{\infty}$ is the free-stream velocity and $h=1$ is the cube height.}
    \label{fig:my_label}
\end{figure}

\begin{figure}
		\centering
		\includegraphics[width=\textwidth]{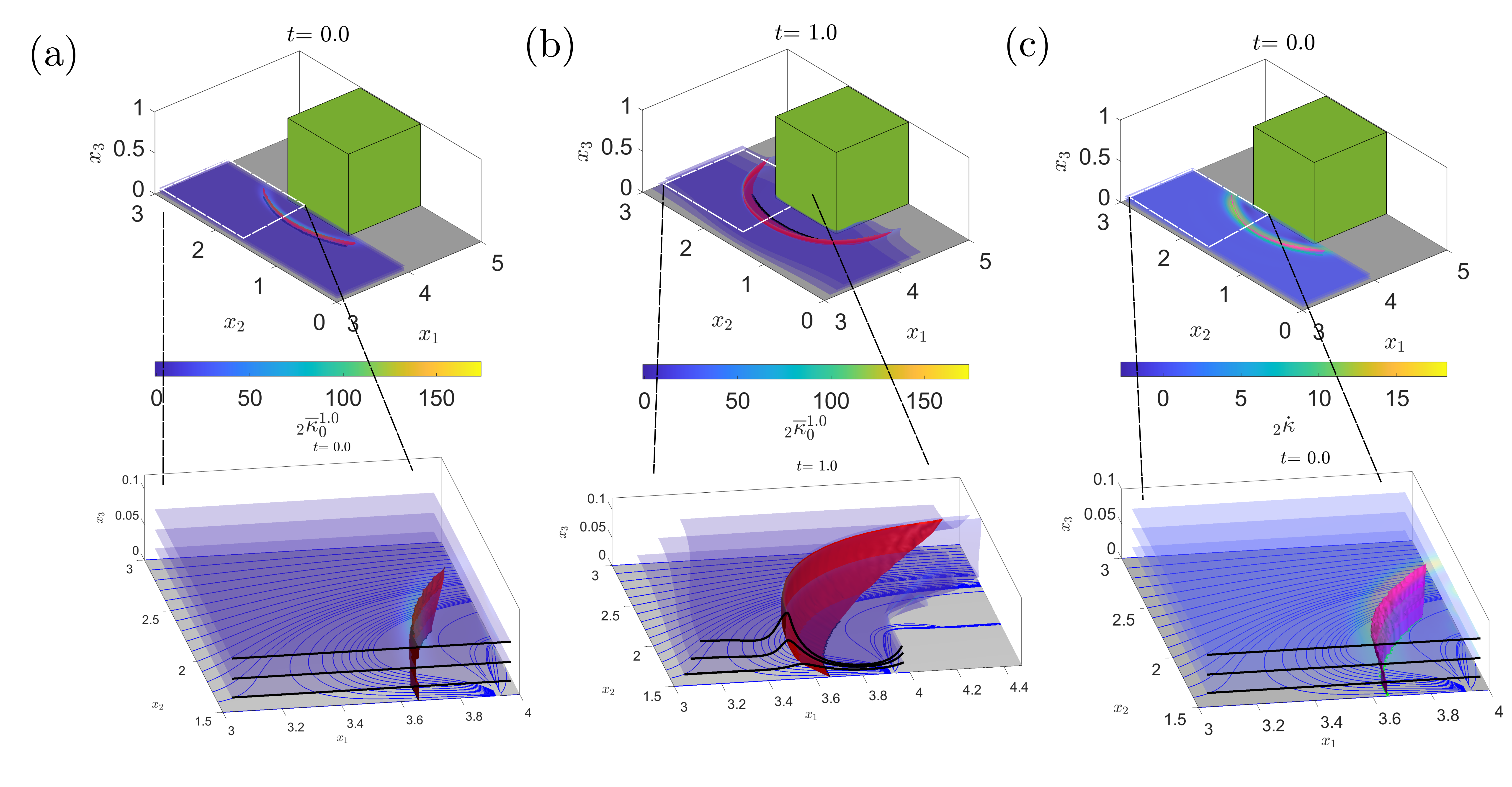}
	\caption{ Separation backbone generated by the flow past a cube described in Section \ref{sec:FlowPastMountedCube}. (a-b) ${}_2\overline{\kappa}_{0}^{1}$ shaded on selected material surfaces at (a) $t=0$ and (b) $t=1$. The Lagrangian backbone of separation $\mathcal{B}(t)$ is in red, and the black line on the wall marks the Lagrangian spiking curve $\gamma_{sc}$. (c) ${}_2\dot{\kappa}$ shaded on selected material surfaces. The Eulerian backbone of separation is in magenta, and the Eulerian spiking curve $\gamma_{scE}$ is in green. Insets show a zoomed view of the above panels and the skin friction lines in blue. Black lines at the edge of material surfaces aid visualization. The time evolution of the above panels, along with the $\bar{H}$ and $\bar{K}$ metrics, are available in \href{https://drive.google.com/file/d/1yIhNvdkAqcDcMRUPmjYsGfZZ3xbDbP-j/view?usp=sharing}{Movie7}.}
		\label{fig:FlowPastCubeP5}
	    \end{figure}

We consider a flow past a mounted cube of edge length $h=1$, placed on the wall at $x_3 =0 $ and centered at $x_1 = 4.5$ and $x_2 = 1.5$. We solve the incompressible Navier--Stokes equations at a Reynolds number $Re = U_{\infty}h/\nu = 200$, where $U_{\infty}$ is the free-stream velocity and $\nu$ the kinematic viscosity. Figure \ref{fig:my_label} shows the flow setup and the velocity streamlines. Additional details regarding the numerical simulation are in Appendix \ref{sec:SteadyFlowCubeDetails}. 
       We explore the separation dynamics in the region upstream of the mounted cube near the wall $x_3 = 0$. Figure \ref{fig:FlowPastCubeP5}a shows the initial position of the Lagrangian backbone of separation $\mathcal{B}(0)$ (red) computed for the time interval $[0,1]$ and the corresponding ${}_2\kappa_{0}^{1}$ field shaded over selected material surfaces at the initial time. Figures \ref{fig:FlowPastCubeP5}a-b show $\mathcal{B}(t)$ along with the ${}_2\kappa_{0}^{t}$ field shaded over material surfaces at $t = 0$ and $t = 1$. Panels (a,b) show again how $\mathcal{B}(t)$ acts as the centerpiece of the separation structure over time. The inset in (a,b) shows the geometry of the separation backbone and how the Lagrangian spiking curve \textcolor{black}{($\gamma_{sc}$)} remains invisible to skin-friction streamlines (blue). Figure \ref{fig:FlowPastCubeP5}c shows the Eulerian backbone of separation $\mathcal{B}_E$ (magenta) and the rate of change of the largest principal curvature ${}_2\dot{\kappa}$ shaded on selected material surfaces. \textcolor{black}{The inset in Figure \ref{fig:FlowPastCubeP5}d shows the Eulerian spiking curve $\gamma_{scE}$(green) and the geometry of $\mathcal{B}_E(0)$. We can see that $\gamma_{scE}\equiv \gamma_{sc}$, which is true for all steady flows (cf.  Tables \ref{tab:sepPointLagra},\ref{tab:sepPointEul}).} As already found in 2D (Fig. \ref{fig:2d_intro} and \citet{serra2018}), the onset of separation is distinct from the on-wall footprint of the corresponding asymptotic structure even in a steady flow.
		
	\subsection{Steady Laminar separation bubble flow}\label{sec:SteadyLSB}
\begin{figure}
    	\includegraphics[width=\textwidth]{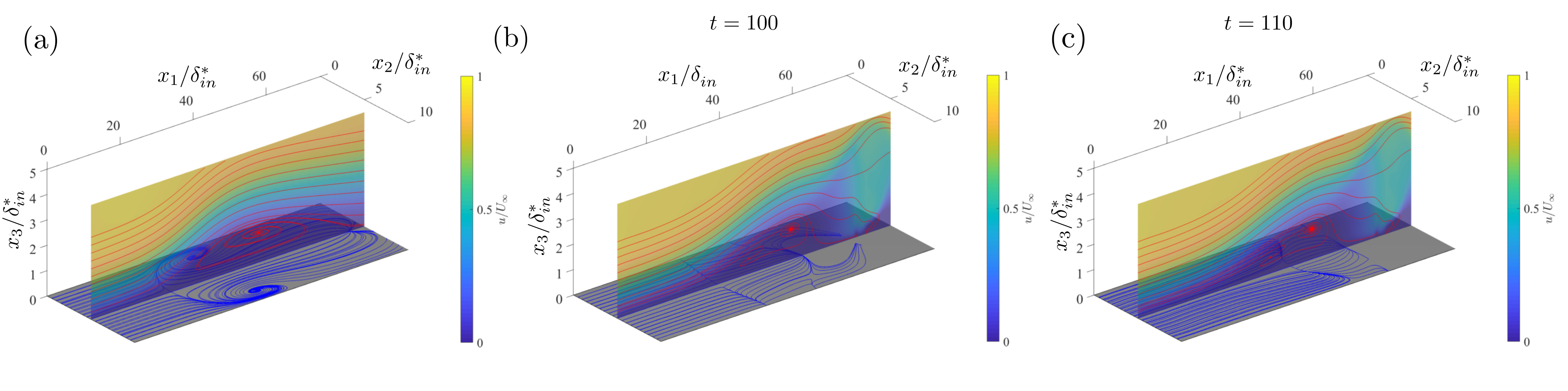}
    	\caption{(a) Flow field of the steady LSB. (b-c) Flow field of the unsteady, moving LSB at (b) $t=100$  and (c) $t=110$. Slice colormap represents the streamwise velocity component $u/U_\infty$, streamlines are in red and skin friction lines are in blue.}
    	\label{fig:LSB_flowField}
\end{figure}

We consider a steady, laminar separation bubble (LSB) on a flat plate with a spanwise modulation. We use an 8th-order accurate discontinuous Galerkin spectral element method \citep{kopriva,Klose2020} to discretize the compressible Navier-Stokes equations spatially. The Reynolds number is $Re_{\delta_{\mathrm{in}}^*} = U_\infty \delta_{\mathrm{in}}^*/\nu = 500$, based on the free-stream velocity $U_\infty$, the height of the inflow boundary layer displacement thickness $\delta_{\mathrm{in}}^*$ and the kinematic viscosity $\nu$.  The free-stream Mach number is 0.3. The computational domain is $L_{x_1}  \times L_{x_2} \times L_{x_3}= 200\delta_{\mathrm{in}}^* \times 10\delta_{\mathrm{in}}^* \times 15\delta_{\mathrm{in}}^*$, where $x_1$, $x_2$ and $x_3$ are the streamwise, spanwise and transverse directions. We prescribe a Blasius profile at the inlet and outlet and set the wall to be isothermal. We prescribe a modified free-stream condition at the top boundary, with a suction profile for lateral velocity component similar to \citet{AlamSandham2000}, to induce flow separation on the bottom wall. Figure \ref{fig:LSB_flowField} (a) shows the streamlines of the flow in red and the skin friction lines in blue. We refer to appendix \ref{App:LSB} for additional details on the setup and the flow field.

\begin{figure}[h]
		\centering
		\includegraphics[width=\textwidth]{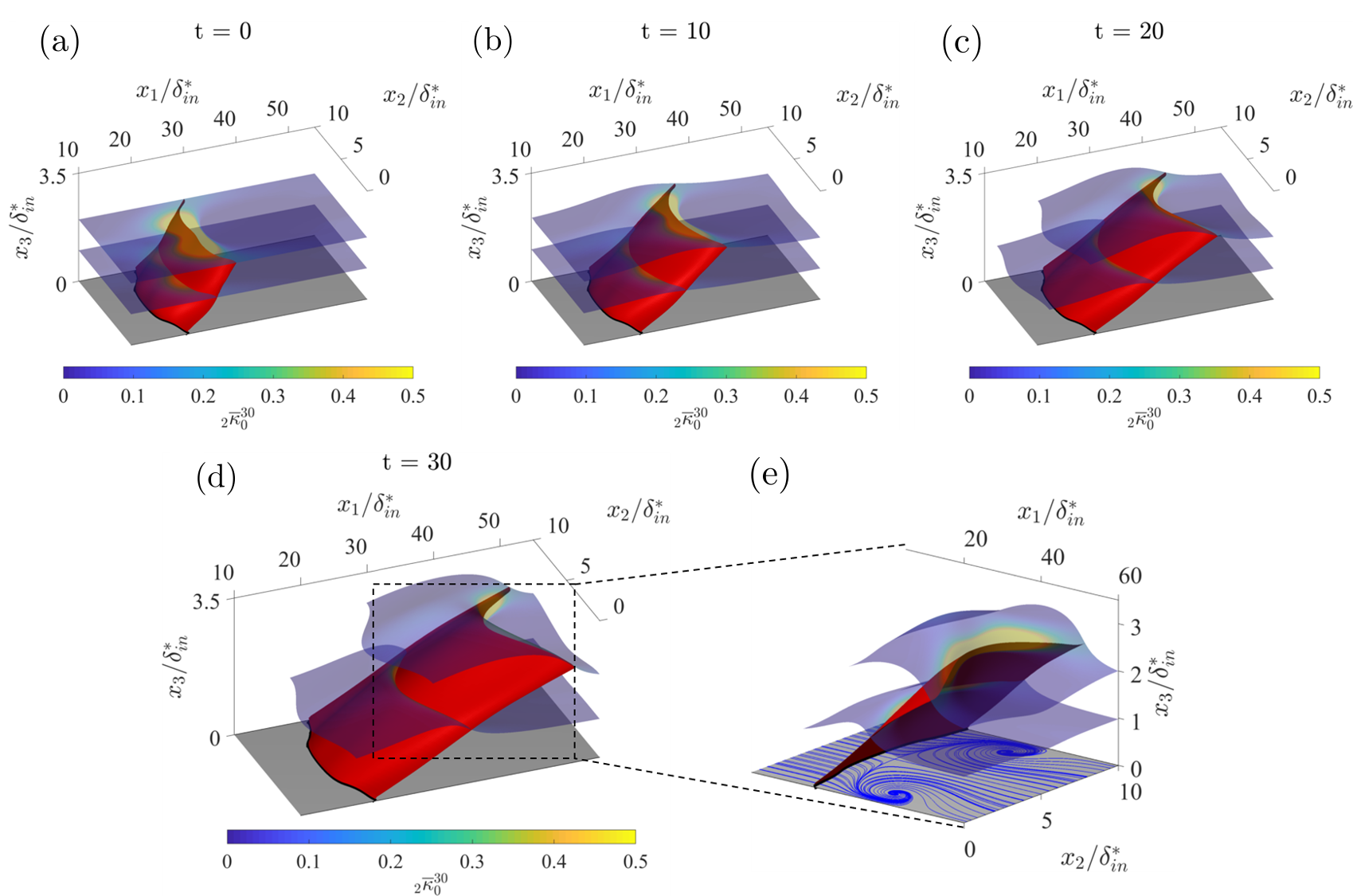}
		\caption{Lagrangian backbone of separation $\mathcal{B}(t)$ (red) generated by a steady LSB on a flat plate with selected material sheets colored by the ${}_2\overline{\kappa}_{0}^{30}$. The Lagrangian spiking curve $\gamma_{sc}$ is in black.
		Panels a-d show advected positions of the backbone $\mathcal{B}(t)$. e) Detailed plot showing the same as (d) from a different view along with skin friction lines in blue. The time evolution of the above panels, along with the $\bar{H}$ and $\bar{K}$ metrics, are available in \href{https://drive.google.com/file/d/14SJtDp5U4mx152rgkUA2eVQrlPcylwkH/view?usp=sharing}{Movie8}.
  \label{fig:steadyLSB}}
    \end{figure}

	\textcolor{black}{
	Figure \ref{fig:steadyLSB} shows selected material sheets colored by the curvature ${}_2\overline{\kappa}_{0}^{30}$. The Lagrangian backbone of separation $\mathcal{B}(t)$ for the time interval $[0,30]$ is in red: its initial position is in panel (a), and its advected positions $\mathcal{B}(t)$ in panels b-d, with a detailed plot in panel e showing skin friction lines in blue. As already noted earlier, the Lagrangian spiking curve (black line) $\gamma_{sc}$ is located upstream of the limiting skin friction line again, indicating that the onset of flow separation has a different location compared to the asymptotic separation structures, even in steady flows.}
    
	\subsection{Unsteady, moving laminar separation bubble flow}\label{sec:MovingLSB}
 
We consider the flow around a LSB on a flat plate with a spanwise modulation introduced in section \ref{sec:SteadyLSB}, but with a time-periodic oscillation of the suction profile on the top boundary to induce unsteady movement of the bubble. The oscillation frequency is $f=1/20$ and the magnitude is $10\delta_{\mathrm{in}}^*$. Figures \ref{fig:LSB_flowField}b-c illustrate the flow's streamlines and skin friction lines. 
 \begin{figure}
		\centering
		\includegraphics[width=\textwidth]{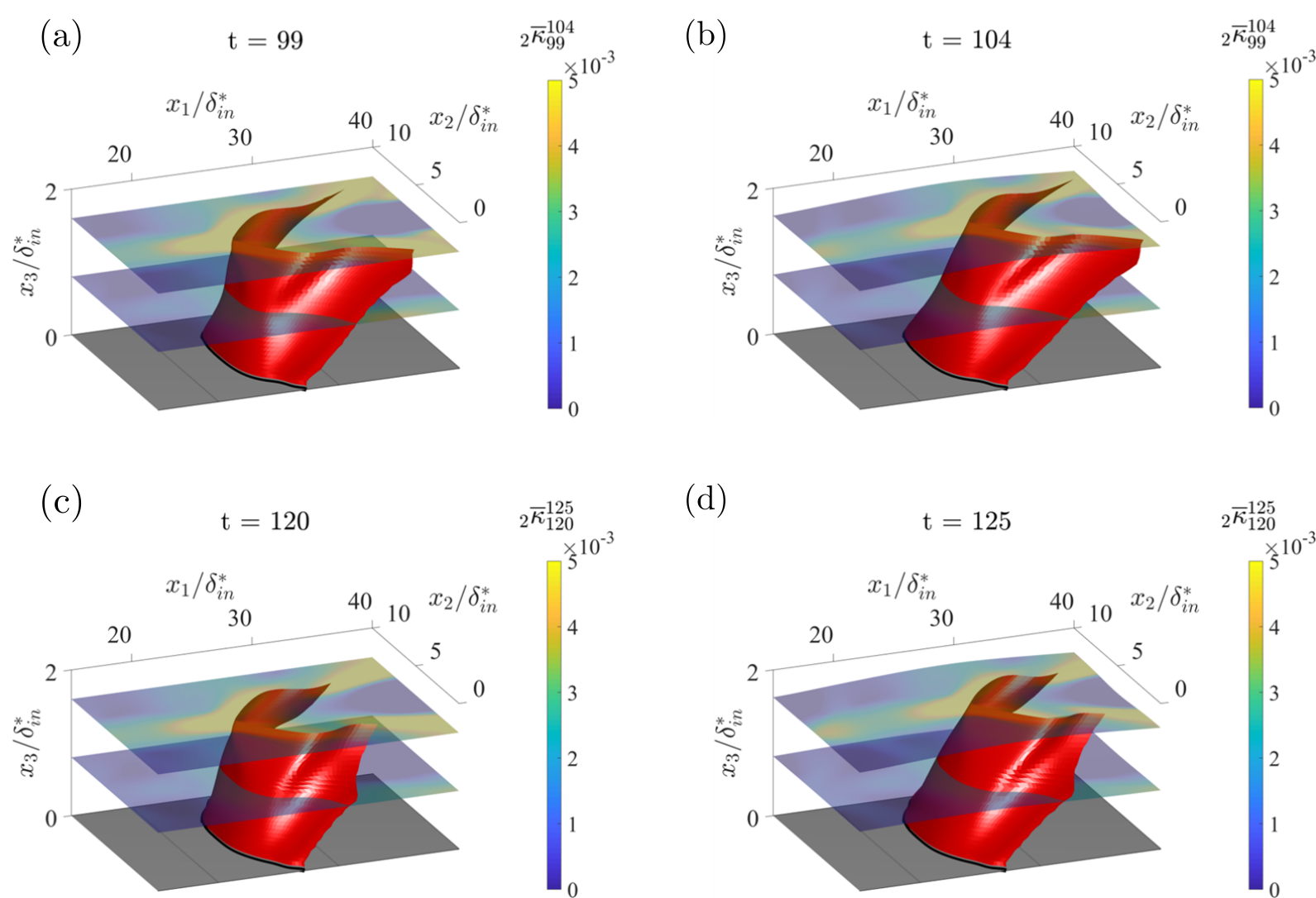}
		\caption{Moving LSB on a flat plate. The lagrangian backbone of separation $\mathcal{B}(t)$ based on ${}_2\overline{\kappa}_{99}^{104}$ (a--b),  ${}_2\overline{\kappa}_{120}^{125}$ (c--d) are in red, along with selected material surfaces colored by the respective ${}_2\overline{\kappa}$ fields, and the Lagrangian spiking curves $\gamma_{sc}$ (black). Panels (a,c) show the material surfaces and $\mathcal{B}(t)$ at the initial times, while panels (b,d) at the final times. The time evolution of the above panels, along with the $\bar{H}$ and $\bar{K}$ metrics, are available in \href{https://drive.google.com/file/d/1hFINeweb0hBpKTs_6jAyoZqceV3nPPMz/view?usp=sharing}{Movie9} for $t\in[99,104]$ and \href{https://drive.google.com/file/d/13J2Fr-P9JNS2uh5PYqf_OquFCoSUKG2D/view?usp=sharing}{Movie10} for $t\in[120,125]$. 
		\label{fig:unsteadyLSB}}
    \end{figure}
We perform a Lagrangian analysis over a time interval of $T=5$ convective time units starting at $t_0$ = 99 (Figs. \ref{fig:unsteadyLSB} a--b) and $t_0$ = 120 (Figs. \ref{fig:unsteadyLSB} c--d), hence compute the Lagrangian backbone of separation based on ${}_2\overline{\kappa}_{99}^{104}$ and ${}_2\overline{\kappa}_{120}^{125}$. 
Figure \ref{fig:unsteadyLSB} highlights the time-dependency of the Lagrangian separation backbone, with its shape and surface signature, i.e., the Lagrangian spiking curve $\gamma_{sc}$, being a function of $t_0$. The backbone and material sheets are shown at their initial positions (a,c) and final positions (b,d).
    \begin{figure}
		\centering
		\includegraphics[width=\textwidth]{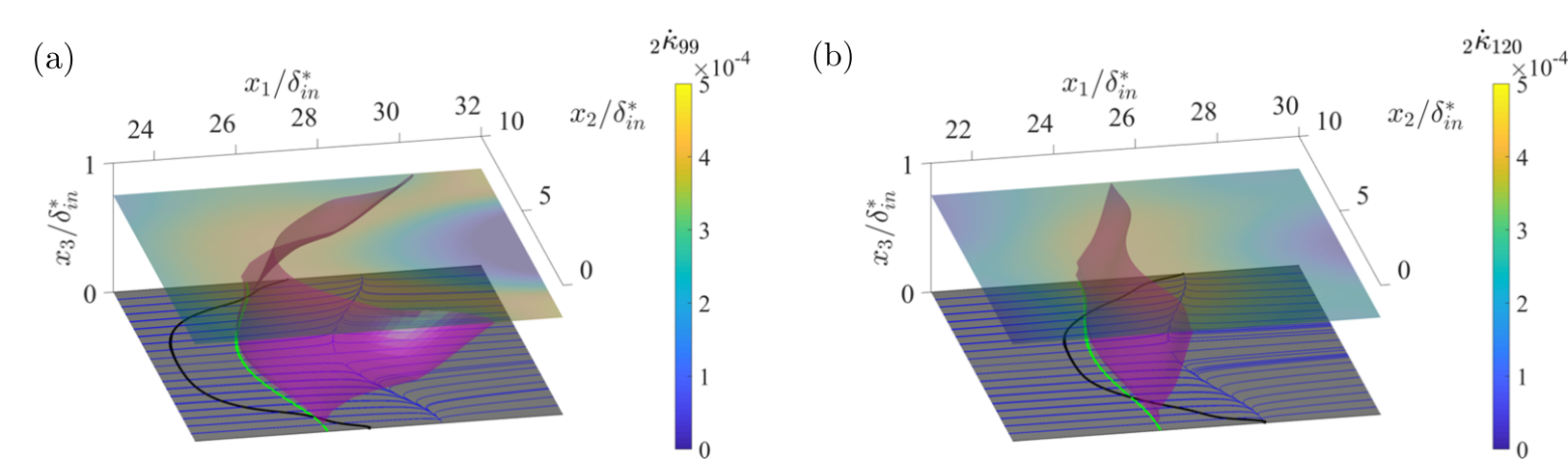}
		\caption{Moving LSB on a flat plate. The Eulerian backbone of separation $\mathcal{B}_E(t)$ is in magenta, the Eulerian spiking curve is in green, and the Lagrangian spiking curves based on $_2\overline{\kappa}_{99}^{104}$ (a) and $_2\overline{\kappa}_{120}^{125}$ (b) are in black.
 Selected material sheets are colored by ${}_2\dot{\kappa}_{99}$ (a) and ${}_2\dot{\kappa}_{120}$. (b) Skin friction lines are in blue.
		\label{fig:unsteadyLSB_Eulerian}}
    \end{figure}

Last, Figure \ref{fig:unsteadyLSB_Eulerian} shows the Eulerian backbone of separation (magenta) at $t_0=99$ and $t_0=120$, their Eulerian spiking curves $\gamma_{scE}$ (green) and the skin friction lines (blue). For comparison, we also show the Lagrangian spiking curves $\gamma_{sc}$ (black) discussed in Fig. \ref{fig:unsteadyLSB}. The Eulerian backbone intersects the wall at a location different from the Lagrangian backbone of separation, consistent with theoretical predictions (cf. Sections \ref{sec:LagrBack}-\ref{sec:EulBack}). Note that the scale of the axes in Figs. \ref{fig:unsteadyLSB_Eulerian} is different to Figs. \ref{fig:unsteadyLSB}. Consistent with the earlier test cases, the backbones of separation we locate remain inaccessible to skin friction lines while acting as the centerpieces of the forming material spikes. 

	\section{Conclusions}
	\label{sec:Conclusions}
We developed a frame-invariant theory of material spike formation during flow separation over a no-slip boundary in three-dimensional flows with any time dependence. Based on the larger principle curvature evolution of material surfaces, our theory uncovers the material spike formation from its birth to its developed Lagrangian structure. Curvature arises from an objective interplay of stretching- and rotation-based kinematic quantities, revealing features that remain hidden to criteria based only on stretching or rotation. Our kinematic theory applies to numerical, experimental or model velocity fields.

The backbone can be one- or two-dimensional, connected to the wall or not. When the backbones connect to the wall, we speak about fixed separation. Otherwise, it is a moving separation. For fixed separation, we have discovered new, distinct, wall locations called spiking points and spiking curves, where one- and two-dimensional backbones connect to the non-slip boundary. We provide criteria for identifying spiking curves and points using wall-based quantities. Remarkably, spiking points and curves remain invisible to classic skin-friction line plots even in steady flows.

Similarly to the spike formation in two dimensions \cite{serra2018}, the spiking points and curves identified here are constant in steady flows and in time-periodic flows analyzed over a time interval that is a multiple of their period. By contrast, they move in general unsteady flows. Our theory is also effective over short-time intervals and admits a rigorous instantaneous limit. These properties, inaccessible to existing criteria, make the present approach promising for monitoring and controlling separation.

The backbone of separation we identify evolves materially under all flow conditions, serving as the core of the separating spike, a universally observed phenomenon unrelated to the flow's time dependence and the presence of singularities in the flow. Two natural next steps are using our results in active flow control strategies and understanding the dynamics or the hydrodynamic forces causing spike formation. 

\section*{Acknowledgements}
GBJ gratefully acknowledges the funding provided by the Air Force Office of Scientific Research under award number FA9550-21-1-0434.

\begin{appendices}
\section{Curvature of a material surface \texorpdfstring{$\mathcal{M}(t)$}{M(t)}}
	\label{sec:differential_geometry}
In this section, we define the quantities to describe the curvature and stretching of material surfaces $\mathcal{M}(t)$. We use the same notation in section \ref{sec:curvature}.
The first fundamental form of $\mathcal{M}(t)$ can be computed as  
\begin{align}
{}_{1}\mathbf{\Gamma}_{t_0}^{t}(\mathbf{p})= \begin{pmatrix}
	\langle \hat{\mathbf{r}}_u, \hat{\mathbf{r}}_u\rangle & \langle \hat{\mathbf{r}}_u, \hat{\mathbf{r}}_v\rangle\\
	\langle \hat{\mathbf{r}}_v, \hat{\mathbf{r}}_u\rangle & \langle \hat{\mathbf{r}}_v, \hat{\mathbf{r}}_v\rangle
	\end{pmatrix}  =\begin{pmatrix} \langle\mathbf{r}_u(\mathbf{p}),C_{t_0}^{t} (\mathbf{r}(\mathbf{p})) \mathbf{r}_u(\mathbf{p})\rangle &\langle\mathbf{r}_u(\mathbf{p}),C_{t_0}^{t} (\mathbf{r}(\mathbf{p})) \mathbf{r}_v(\mathbf{p})\rangle \\  \langle\mathbf{r}_u(\mathbf{p}),C_{t_0}^{t} (\mathbf{r}(\mathbf{p})) \mathbf{r}_v(\mathbf{p})\rangle&\langle\mathbf{r}_v(\mathbf{p}),C_{t_0}^{t} (\mathbf{r}(\mathbf{p})) \mathbf{r}_v(\mathbf{p})\rangle  \end{pmatrix}.
\end{align} 
The second fundamental form of $\mathcal{M}(t)$ is given by
\begin{align}\label{eqn:DefineSecondForm}
{}_{2}\mathbf{\Gamma}_{t_0}^{t}(\mathbf{p}) &= \begin{pmatrix} \langle\mathbf{n}_t, \hat{\mathbf{r}}_{uu}(\mathbf{p})\rangle &\langle\mathbf{n}_t, \hat{\mathbf{r}}_{uv}(\mathbf{p})\rangle \\  \langle\mathbf{n}_t, \hat{\mathbf{r}}_{uv}(\mathbf{p})\rangle&\langle\mathbf{n}_t, \hat{\mathbf{r}}_{vv}(\mathbf{p})\rangle  \end{pmatrix}\notag \\
&= \begin{pmatrix} \langle\mathbf{n}_t, \mathbf{\nabla}^2\mathbf{F}^{t}_{t_0}(\mathbf{r}(\mathbf{p}))\mathbf{r}_u\mathbf{r}_u+\mathbf{\nabla}\mathbf{F}_{t_0}^{t}(\mathbf{r}(\mathbf{p}))\mathbf{r}_{uu}\rangle &\langle\mathbf{n}_t, \mathbf{\nabla}^2\mathbf{F}^{t}_{t_0}(\mathbf{r}(\mathbf{p}))\mathbf{r}_u\mathbf{r}_v+\mathbf{\nabla}\mathbf{F}_{t_0}^{t}(\mathbf{r}(\mathbf{p}))\mathbf{r}_{uv}\rangle \\  \langle\mathbf{n}_t, \mathbf{\nabla}^2\mathbf{F}^{t}_{t_0}(\mathbf{r}(\mathbf{p}))\mathbf{r}_v\mathbf{r}_u+\mathbf{\nabla}\mathbf{F}_{t_0}^{t}(\mathbf{r}(\mathbf{p}))\mathbf{r}_{vu}\rangle&\langle\mathbf{n}_t, \mathbf{\nabla}^2\mathbf{F}^{t}_{t_0}(\mathbf{r}(\mathbf{p}))\mathbf{r}_v\mathbf{r}_v+\mathbf{\nabla}\mathbf{F}_{t_0}^{t}(\mathbf{r}(\mathbf{p}))\mathbf{r}_{vv}\rangle  \end{pmatrix}\notag\\
& = \mathbf{B}_{t_0}^{t}(\mathbf{p})+\mathbf{A}_{t_0}^{t}(\mathbf{p}),
\end{align}
where $\textcolor{black}{[\mathbf{\nabla}^2\mathbf{F}^t_{t_0}(\mathbf{r}(\mathbf{p})\mathbf{r}_u\mathbf{r}_v]_i = [\mathbf{\nabla}^2\mathbf{F}^t_{t_0}(\mathbf{r})\mathbf{r}_{u}]_{ij}  [\mathbf{r}_{v}]_{j}= \langle \mathbf{\nabla}[\mathbf{\nabla}\mathbf{F}^t_{t_0}(\mathbf{r})]_{ij},\mathbf{r}_{u}\rangle[\mathbf{r}_{v}]_{j}}$\footnote[2]{$[\mathbf{\nabla}^2\mathbf{F}^t_{t_0}(\mathbf{r})\mathbf{r}_{u}]_{ij}$ represents the directional derivatives of $[\mathbf{\nabla}\mathbf{F}^t_{t_0}(\mathbf{r})]_{ij}$ in the direction $\mathbf{r}_{u}$. We use the same notation in eqs. \eqref{eq:Bdott0}-\eqref{eq:nabla2vDecomp}.} and
	\begin{align}
	    \mathbf{B}_{t_0}^{t}(\mathbf{p}) &= \begin{pmatrix}
	\langle \mathbf{n}_t,\mathbf{\nabla}^2\mathbf{F}^{t}_{t_0}(\mathbf{r}(\mathbf{p}))\mathbf{r}_u\mathbf{r}_u\rangle & \langle \mathbf{n}_t,\mathbf{\nabla}^2\mathbf{F}^{t}_{t_0}(\mathbf{r}(\mathbf{p}))\mathbf{r}_u\mathbf{r}_v\rangle \\
	\langle \mathbf{n}_t,\mathbf{\nabla}^2\mathbf{F}^{t}_{t_0}(\mathbf{r}(\mathbf{p}))\mathbf{r}_v\mathbf{r}_u\rangle&\langle \mathbf{n}_t,\mathbf{\nabla}^2\mathbf{F}^{t}_{t_0}(\mathbf{r}(\mathbf{p}))\mathbf{r}_v\mathbf{r}_v\rangle
	\end{pmatrix},\label{eq:Btt0}\\
	\mathbf{A}_{t_0}^{t}(\mathbf{p}) &= \begin{pmatrix}
	\langle \mathbf{n}_t,\mathbf{\nabla}\mathbf{F}_{t_0}^{t}(\mathbf{r}(\mathbf{p}))\mathbf{r}_{uu}\rangle & \langle \mathbf{n}_t,\mathbf{\nabla}\mathbf{F}_{t_0}^{t}(\mathbf{r}(\mathbf{p}))\mathbf{r}_{uv}\rangle \\
	\langle \mathbf{n}_t,\mathbf{\nabla}\mathbf{F}_{t_0}^{t}(\mathbf{r}(\mathbf{p}))\mathbf{r}_{vu}\rangle&\langle \mathbf{n}_t,\mathbf{\nabla}\mathbf{F}_{t_0}^{t}(\mathbf{r}(\mathbf{p}))\mathbf{r}_{vv}\rangle 
	\end{pmatrix}\label{eq:Att0}
	\end{align}
	The Weingarten map of $\mathcal{M}(t)$ is given by
	\begin{align}\label{eq:DefWeingarten}
	\mathbf{W}_{t_0}^{t}(\mathbf{p}) = ({}_1\mathbf{\Gamma}(\mathbf{p}))^{-1}{}_2\mathbf{\Gamma}_{t_0}^{t}(\mathbf{p})	.
	\end{align}
The principle curvatures ${}_{1}\kappa_{t_0}^t(\mathbf{p},t)$ and ${}_{2}\kappa_{t_0}^t(\mathbf{p},t)$ are given by the eigenvalues of the Weingarten map $\mathbf{W}_{t_0}^{t}(\mathbf{p})$.
\section{Proof of Theorem \ref{theorem1}}
		
Here we derive the Lagrangian evolution of the Weingarten map along a material surface $\mathcal{M}(t)$. 
		\label{sec:appendixB}
		\subsection{Material Evolution of the Weingarten Map}
		
		We expand \eqref{eq:Att0} as
		\begin{align}
		\mathbf{A}_{t_0}^{t}(\mathbf{p}) &= \frac{1}{J_{t_0}^{t}(\mathbf{p})}\left(\begin{smallmatrix}
		\langle(\mathbf{\nabla}\mathbf{F}^t_{t_0}\mathbf{r}_u)\times(\mathbf{\nabla}\mathbf{F}^t_{t_0}\mathbf{r}_v),\mathbf{\nabla}\mathbf{F}^t_{t_0}\mathbf{r}_{uu}\rangle&\langle(\mathbf{\nabla}\mathbf{F}^t_{t_0}\mathbf{r}_u)\times(\mathbf{\nabla}\mathbf{F}^t_{t_0}\mathbf{r}_v),\mathbf{\nabla}\mathbf{F}^t_{t_0}\mathbf{r}_{uv}\rangle\\
		\langle(\mathbf{\nabla}\mathbf{F}^t_{t_0}\mathbf{r}_u)\times(\mathbf{\nabla}\mathbf{F}^t_{t_0}\mathbf{r}_v),\mathbf{\nabla}\mathbf{F}^t_{t_0}\mathbf{r}_{vu}\rangle&\langle(\mathbf{\nabla}\mathbf{F}^t_{t_0}\mathbf{r}_u)\times(\mathbf{\nabla}\mathbf{F}^t_{t_0}\mathbf{r}_v),\mathbf{\nabla}\mathbf{F}^t_{t_0}\mathbf{r}_{vv}\rangle
		\end{smallmatrix}\right),\label{eqn:Att0expand}
		\end{align}
		where $\mathbf{\nabla}\mathbf{F}^t_{t_0} = \mathbf{\nabla}\mathbf{F}^t_{t_0}(\mathbf{r}(\mathbf{p}))$ and $J_{t_0}^{t}(\mathbf{p}) = \sqrt{det({}_1\mathbf{\Gamma}(\mathbf{p})_{t_0}^{t})}$. We simplify \eqref{eqn:Att0expand} as
		\begin{align}
		\mathbf{A}_{t_0}^{t}(\mathbf{p}) &= \frac{1}{J_{t_0}^{t}(\mathbf{p})}\left(\begin{smallmatrix}
\langle(\mathbf{\nabla}\mathbf{F}^t_{t_0}\mathbf{r}_u)\times(\mathbf{\nabla}\mathbf{F}^t_{t_0}\mathbf{r}_v),\mathbf{\nabla}\mathbf{F}^t_{t_0}\mathbf{r}_{uu}\rangle&\langle(\mathbf{\nabla}\mathbf{F}^t_{t_0}\mathbf{r}_u)\times(\mathbf{\nabla}\mathbf{F}^t_{t_0}\mathbf{r}_v),\mathbf{\nabla}\mathbf{F}^t_{t_0}\mathbf{r}_{uv}\rangle \nonumber\\
		\langle(\mathbf{\nabla}\mathbf{F}^t_{t_0}\mathbf{r}_u)\times(\mathbf{\nabla}\mathbf{F}^t_{t_0}\mathbf{r}_v),\mathbf{\nabla}\mathbf{F}^t_{t_0}\mathbf{r}_{vu}\rangle&\langle(\mathbf{\nabla}\mathbf{F}^t_{t_0}\mathbf{r}_u)\times(\mathbf{\nabla}\mathbf{F}^t_{t_0}\mathbf{r}_v),\mathbf{\nabla}\mathbf{F}^t_{t_0}\mathbf{r}_{vv}\rangle
		\end{smallmatrix}\right) \nonumber\\
		&= \frac{\textup{det}(\mathbf{\nabla}\mathbf{F}^t_{t_0}(\mathbf{r}(\mathbf{p})))}{J_{t_0}^{t}(\mathbf{p})}\begin{pmatrix}
		\langle\mathbf{r}_u\times \mathbf{r}_v, \mathbf{r}_{uu}\rangle &\langle\mathbf{r}_u\times \mathbf{r}_v, \mathbf{r}_{uv}\rangle\nonumber \\
		\langle\mathbf{r}_u\times\mathbf{r}_v, \mathbf{r}_{vu}\rangle &\langle\mathbf{r}_u\times \mathbf{r}_v,\mathbf{r}_{vv}\rangle
		\end{pmatrix}\nonumber\\
		& =\frac{J_{t_0}(\mathbf{p})\textup{det}(\mathbf{\nabla}\mathbf{F}^t_{t_0}(\mathbf{r}(\mathbf{p})))}{J_{t_0}^{t}(\mathbf{p})}\frac{1}{J_{t_0}(\mathbf{p})}\begin{pmatrix}
		\langle\mathbf{r}_u\times \mathbf{r}_v,\mathbf{r}_{uu}\rangle&\langle\mathbf{r}_u\times \mathbf{r}_v,\mathbf{r}_{uv}\rangle\\
		\langle\mathbf{r}_u\times \mathbf{r}_v, \mathbf{r}_{vu}\rangle&\langle\mathbf{r}_u\times \mathbf{r}_v, \mathbf{r}_{vv}\rangle\nonumber\\
		\end{pmatrix}\\
		&= \frac{J_{t_0}(\mathbf{p})\textup{det}(\mathbf{\nabla}\mathbf{F}^t_{t_0}(\mathbf{r}(\mathbf{p})))}{J_{t_0}^{t}(\mathbf{p})}\mathbf{A}_{t_0}^{t_0}(\mathbf{p}).
		\end{align}
		Because $\mathbf{B}_{t_0}^{t_0}(\mathbf{p})$ = $\mathbf{0}$, using \eqref{eqn:DefineSecondForm}, we have ${}_2\mathbf{\Gamma}_{t_0}(\mathbf{p}) = \mathbf{A}_{t_0}^{t_0}(\mathbf{p})$. Therefore ${}_2\mathbf{\Gamma}_{t_0}(\mathbf{p})$ is given by 
		\begin{equation}\label{eq:2forminit}
		{}_2\mathbf{\Gamma}_{t_0}^{t}(\mathbf{p}) = \frac{J_{t_0}(\mathbf{p})\textup{det}(\mathbf{\nabla}\mathbf{F}^t_{t_0}(\mathbf{r}(\mathbf{p})))}{J_{t_0}^{t}(\mathbf{p})}{}_2\mathbf{\Gamma}_{t_0}(\mathbf{p})+ \mathbf{B}_{t_0}^{t}(\mathbf{p}).
		\end{equation}
		By substituting \eqref{eq:2forminit} into the definition of the Weingarten map \eqref{eq:DefWeingarten}, we obtain 
		\begin{equation}\label{eq:weingartenComplete}
		\mathbf{W}_{t_0}^{t}(\mathbf{p}) = ({}_1\mathbf{\Gamma}_{t_0}^{t}(\mathbf{p}))^{-1}{}_1\mathbf{\Gamma}_{t_0}(\mathbf{p})\frac{J_{t_0}(\mathbf{p}) \textup{det}(\mathbf{\nabla}\mathbf{F}^t_{t_0}(\mathbf{r}(\mathbf{p})))}{J_{t_0}^{t}(\mathbf{p})}\mathbf{W}_{t_0}(\mathbf{p})+({}_1\mathbf{\Gamma}_{t_0}^{t}(\mathbf{p}))^{-1}\mathbf{B}_{t_0}^{t}(\mathbf{p}).
		\end{equation}

	\subsection{Rate of change of the Weingarten map of a material surface}
	We take the total time derivative $\frac{d}{dt}(\cdot):= \dot{(\cdot)} $ of \eqref{eq:weingartenComplete} and evaluate it at $t =t_0$. For clarity, we calculate the time derivatives of each term separately:
	\begin{align}
	    	\dot{\overline{\text{det}(\mathbf{\nabla}\mathbf{F}_{t_0}^{t} (\mathbf{r}(\mathbf{p}))}}\vert_{t_0} &=\mathbf{\nabla} \cdot\mathbf{f}(\mathbf{r}(\mathbf{p}),t_0),\label{eq:Cdott0}\\
	        \dot{\overline{J_{t_0}^{t}(\mathbf{p})}\vert_{t_0}} &= \frac{1}{J_{t_0}(\mathbf{p})}(\langle\mathbf{r}_{u},\mathbf{S}(\mathbf{r}(\mathbf{p}),t_0)\mathbf{r}_{u}\rangle\langle\mathbf{r}_{v},\mathbf{r}_{v}\rangle+\langle\mathbf{r}_{v},\mathbf{S}(\mathbf{r}(\mathbf{p}),t_0)\mathbf{r}_{v}\rangle\langle\mathbf{r}_{u},\mathbf{r}_{u}\rangle\notag,\\ &\quad \ \ \ -2\langle\mathbf{r}_{u},\mathbf{S}(\mathbf{r}(\mathbf{p}),t_0)\mathbf{r}_{v}\rangle\langle\mathbf{r}_{u},\mathbf{r}_{v}\rangle)\notag\\ &= \frac{\alpha_{t_0}}{J_{t_0}(\mathbf{p})},
	\end{align}
	
	where $\alpha_{t_0} = \langle\mathbf{r}_{u},\mathbf{S}(\mathbf{r}(\mathbf{p}))\mathbf{r}_{u}\rangle\langle\mathbf{r}_{v},\mathbf{r}_{v}\rangle+\langle\mathbf{r}_{v},\mathbf{S}(\mathbf{r}(\mathbf{p}),t_0)\mathbf{r}_{v}\rangle\langle\mathbf{r}_{u},\mathbf{r}_{u}\rangle-2\langle\mathbf{r}_{u},\mathbf{S}(\mathbf{r}(\mathbf{p}),t_0)\mathbf{r}_{v}\rangle\langle\mathbf{r}_{u},\mathbf{r}_{v}\rangle$,
	\begin{align}
	    \dot{\overline{\mathbf{C}^t_{t_0}}}\vert_{t_0}&= 2\mathbf{S}(\mathbf{r}(\mathbf{p}),t_0),\\
	    	\dot{\overline{\mathbf{B}_{t_0}^{t}}}\vert_{t_0} &= \begin{pmatrix}
	\langle\mathbf{n}_{t_0},\nabla^2\mathbf{f}(\mathbf{r}(\mathbf{p}),t_0)\mathbf{r}_{u}\mathbf{r}_{u}\rangle&\langle\mathbf{n}_{t_0},\nabla^2\mathbf{f}(\mathbf{r}(\mathbf{p}),t_0)\mathbf{r}_{u}\mathbf{r}_{v}\rangle,\\
	\langle\mathbf{n}_{t_0},\nabla^2\mathbf{f}(\mathbf{r}(\mathbf{p}),t_0)\mathbf{r}_{v}\mathbf{r}_{u}\rangle&\langle\mathbf{n}_{t_0},\nabla^2\mathbf{f}(\mathbf{r}(\mathbf{p}),t_0)\mathbf{r}_{v}\mathbf{r}_{v}\rangle
	\end{pmatrix}\label{eq:Bdott0}\\
	\dot{\overline{({}_1\mathbf{\Gamma}_{t_0}^{t}(\mathbf{p})^{-1})}}\vert_{t_0} &= -\frac{2\alpha_{t_0}}{J_{t_0}^2(\mathbf{p})}({}_1\mathbf{\Gamma}_{t_0}(\mathbf{p}))^{-1}+\frac{2}{J_{t_0}^2(\mathbf{p})}\mathbf{D}(\mathbf{p},t_0),\\ &\text{$\quad$ }\mathbf{D}(\mathbf{p}, t_0) = \left(\begin{smallmatrix}
		\langle\mathbf{r}_v,\mathbf{S}(\mathbf{r}(\mathbf{p}),t_0)\mathbf{r}_v\rangle&-\langle\mathbf{r}_u,\mathbf{S}(\mathbf{r}(\mathbf{p}),t_0)\mathbf{r}_v\rangle\\
		-\langle\mathbf{r}_u,\mathbf{S}(\mathbf{r}(\mathbf{p}),t_0)\mathbf{r}_v\rangle&\langle\mathbf{r}_u,\mathbf{S}(\mathbf{r}(\mathbf{p}),t_0)\mathbf{r}_u\rangle
		\end{smallmatrix}\right).
	\end{align}

	We can rewrite $\dot{\overline{\mathbf{B}}}\vert_{t_0}$ in terms of the gradients of the rate-of-strain tensor and vorticity as
	\begin{equation}\label{eq:nabla2vDecomp}
	\mathbf{\nabla}^2\mathbf{f}(\mathbf{r}(\mathbf{p}),t_0) = \mathbf{\nabla}\mathbf{S}(\mathbf{r}(\mathbf{p}),t_0)+\mathbf{\nabla}\mathbf{\Omega}(\mathbf{r}(\mathbf{p}),t_0).
	\end{equation}
	Using \eqref{eq:nabla2vDecomp} and \eqref{eq:Bdott0}, we get 
	\begin{align}\label{eq:Bdott0expand}
	\dot{\overline{\mathbf{B}_{t_0}^{t}}}\vert_{t_0} = \underbrace{\left(\begin{smallmatrix}
	\langle\mathbf{n}_{t_0},\nabla\mathbf{S}(\mathbf{r}(\mathbf{p}),t_0)\mathbf{r}_{u}\mathbf{r}_{u}\rangle&\langle\mathbf{n}_{t_0},\nabla\mathbf{S}(\mathbf{r}(\mathbf{p}),t_0)\mathbf{r}_{u}\mathbf{r}_{v}\rangle\\
	\langle\mathbf{n}_{t_0},\nabla\mathbf{S}(\mathbf{r}(\mathbf{p}),t_0)\mathbf{r}_{v}\mathbf{r}_{u}\rangle&\langle\mathbf{n}_{t_0},\nabla\mathbf{S}(\mathbf{r}(\mathbf{p}),t_0)\mathbf{r}_{v}\mathbf{r}_{v}\rangle\end{smallmatrix}\right)}_{\mathbf{M}(\mathbf{p},t_0)} + \underbrace{\left(\begin{smallmatrix}
	\langle\mathbf{n}_{t_0},\nabla\mathbf{\Omega}(\mathbf{r}(\mathbf{p}),t_0)\mathbf{r}_{u}\mathbf{r}_{u}\rangle&\langle\mathbf{n}_{t_0},\nabla\mathbf{\Omega}(\mathbf{r}(\mathbf{p}),t_0)\mathbf{r}_{u}\mathbf{r}_{v}\rangle\\
	\langle\mathbf{n}_{t_0},\nabla\mathbf{\Omega}(\mathbf{r}(\mathbf{p}),t_0)\mathbf{r}_{v}\mathbf{r}_{u}\rangle&\langle\mathbf{n}_{t_0},\nabla\mathbf{\Omega}(\mathbf{r}(\mathbf{p}),t_0)\mathbf{r}_{v}\mathbf{r}_{v}\rangle\end{smallmatrix}\right)}_{\mathbf{N}(\mathbf{p},t_0)}.
	\end{align}
Using (\ref{eq:Cdott0}-\ref{eq:Bdott0expand}), we calculate the time derivative of the material evolution of the Weingarten map \eqref{eq:weingartenComplete} at $t=t_0$, which is given by,
        \begin{multline}
		\dot{\mathbf{W}}_{t_0}(\mathbf{p})= \underbrace{\left[\left( \mathbf{\nabla}\cdot\mathbf{f}(\mathbf{r}(\mathbf{p}),t_0)-\frac{3\alpha_{t_0}(\mathbf{p})}{J_{t_0}^2(\mathbf{p})}\right)\mathbf{I}+\frac{2\mathbf{D}(\mathbf{p},t_0){}_1\mathbf{\Gamma}_{t_0}(\mathbf{p})}{J_{t_0}^2(\mathbf{p})}\right]\mathbf{W}_{t_0}}_{\dot{\mathbf{W}}_{I}}\\+\underbrace{({}_1\mathbf{\Gamma}_{t_0}(\mathbf{p}))^{-1}\mathbf{M}(\mathbf{p},t_0)}_{{\dot{\mathbf{W}}_{II}}}+\underbrace{({}_1\mathbf{\Gamma}_{t_0}(\mathbf{p                                     }))^{-1}\mathbf{N}_{t_0}}_{\dot{\mathbf{W}}_{III}}.
		\end{multline}
\section{ \texorpdfstring{$\mathbf{W}_{t_0}^t(\mathbf{p})$}{Wt0} is invariant under changes of parametrization and Euclidean coordinate transformations}\label{App:InvarianceProof}
Here we show that the folding of a material surface $\mathbf{W}_{t_0}^t(\mathbf{p})$ (cf. eq. \eqref{eq:Wlagrangian}) is independent of parametrization, i.e. the choice of $\mathbf{r}(\mathbf{p})$ (cf. eq. \eqref{eq:M(t)}), as well as of Euclidean coordinate changes of the form 
\begin{equation*}
\tilde{\mathbf{x}}=\mathbf{Q}(t)\mathbf{x}+\mathbf{b}(t),
\end{equation*}
where $\mathbf{Q}(t)\in SO(3)$ and $\mathbf{b}(t)\in\mathbb{R}^{3}$ are smooth functions of time. The invariance of $\mathbf{W}_{t_0}^t(\mathbf{p})$ implies that $\tilde{\mathbf{W}}_{t_0}^t(\mathbf{p})=\mathbf{W}_{t_0}^t(\mathbf{p})$, 
where $\tilde{(\cdot)}$ denotes quantities expressed as a function of the new $\tilde{\mathbf{x}}-$coordinate, and $(\cdot)$ the same quantity expressed in terms of the original $\mathbf{x}-$coordinate. We note that this is a stronger property than objectivity \citep{TruesdellNoll2004}. 
To show this invariance, it suffices to note that $\mathbf{W}_{t_0}^t(\mathbf{p})$ is the Weingarten map of a surface $\mathcal{M}(t)$ parametrized by $\hat{\mathbf{r}}_{t_0}^t(\mathbf{p})=\mathbf{F}_{t_0}^t[\mathbf{r}(\mathbf{p})]$ (eq. \eqref{eq:M(t)} and Fig. \ref{fig:intro}a), and recall that the Weingarten map is independent of the parametrization $\hat{\mathbf{r}}_{t_0}^t(\mathbf{p})$ \citep{Kuhnel2015}. This property still holds in our context where $\hat{\mathbf{r}}_{t_0}^t(\mathbf{p})$ is a composition of the parametrization of the initial surface$\mathcal{M}(t_0)$ and the action of $\mathbf{F}_{t_0}^t$, which is affected by \eqref{eq:CoordchangeObj}. This completes the proof of Proposition \ref{sec:Prop1}.

\section{Lagrangian spiking points and curves}\label{App:LagrSpikingPoint}
Here we derive the analytical expressions for the Lagrangian spiking point $\mathbf{r}_{sp} = \mathbf{r}(\mathbf{p}_{sp})$ and Lagrangian spiking curve $\gamma_{sc} = \mathbf{r}(\mathbf{p}_{sc})$, i.e. where the Lagrangian backbone of separation $\mathcal{B}(t_0)$ connects with the wall. 
\subsection{Compressible flows}\label{App:LagrSpikingPointCompr}
Because of the no-slip condition, the wall is invariant, which implies 
\begin{equation}\label{eq:InvariantWall}
    \overline{\mathbf{W}}_{t_0}^{t}(\mathbf{r_{\eta=0}}(\mathbf{p})) = \mathbf{0}.
\end{equation}
Therefore, to identify $\mathbf{p}_{sp}$ and $\mathbf{p}_{sc}$, we derive the Weingarten map infinitesimally close to the wall $\overline{\mathbf{W}}_{t_0}^{t}(\mathbf{r_{\eta=\delta\eta}}(\mathbf{p}))$ by Taylor expanding $\overline{\mathbf{W}}_{t_0}^{t}(\mathbf{r_{\eta}}(\mathbf{p}))$ along $\eta$ and using \eqref{eq:InvariantWall}, which gives 
\begin{equation}\label{eq:deltaEtaLagr}
    \overline{\mathbf{W}}_{t_0}^{t}(\mathbf{r_{\delta\eta}}(\mathbf{p})) = \underbrace{\partial_{\eta}\overline{\mathbf{W}}_{t_0}^{t}(\mathbf{r_{\eta}}(\mathbf{p}))\vert_{\eta=0}}_{\partial_{\eta}\overline{\mathbf{W}}_{\eta}(\mathbf{p})\vert_{\eta=0}}\delta\eta + O(\delta\eta^2).
\end{equation}
Using $\partial_{\eta}\overline{\mathbf{W}}_{\eta}(\mathbf{p})\vert_{\eta=0}$, we calculate  eigenvalues and eigenvectors of $\overline{\mathbf{W}}_{t_0}^{t}(\mathbf{r_{\delta\eta}}(\mathbf{p}))$ to the leading order in $\eta$. Therefore, we use \eqref{eq:deltaEtaLagr} and the criteria described in Proposition \ref{sec:Prop2} to determine $\mathbf{p}_{sp}$ and $\mathbf{p}_{sc}$.

To gain further insight into $\partial_{\eta}\overline{\mathbf{W}}_{\eta}(\mathbf{p})\vert_{\eta=0}$ we express it in terms of the spatial derivatives of Eulerian quantities
\begin{equation}
    \partial_{\eta}\overline{\mathbf{W}}_{t_0}^{t}(\mathbf{r_{\eta}}(\mathbf{p}))\vert_{\eta=0} = \partial_{\eta}\int_{t_0}^{t} \dot{\overline{\mathbf{W}}}_{t_0}^{\tau}(\mathbf{p})\vert_{\eta=0}d\tau = \int_{t_0}^{t} \partial_{\eta}\dot{\overline{\mathbf{W}}}_{t_0}^{\tau}(\mathbf{p})\vert_{\eta=0}d\tau,
\end{equation}
where $\dot{\overline{\mathbf{W}}}_{t_0}^{t}(\mathbf{p})=\dot{\mathbf{W}}_{t_0}^{t}(\mathbf{p})$ (cf \eqref{eq:Wchange}) is evaluated along trajectories of \eqref{eq:VelODE}. Because of the no-slip condition on the wall, the convective term in the material derivative of $\dot{\overline{\mathbf{W}}}_{t_0}^{t}(\mathbf{p})$ is identically zero at $\eta=0$. Assuming a flat no-slip wall and using \eqref{eq:InvariantWall} we have,

\begin{equation}\label{eq:InvariantFlatWall}
    \mathbf{W}_{t_0}^{t}(\mathbf{r_{\eta=0}}(\mathbf{p})) = \mathbf{0}. 
\end{equation}
Since the wall is a flat invariant set, ${}_{1}\mathbf{\Gamma}_{t_0}^{t}(\mathbf{p})\vert_{\eta=0} = \mathbf{I}$. Using \eqref{eq:InvariantFlatWall} and \eqref{eq:weingartenComplete} we have,
\begin{equation}\label{eq:Bzero}
    \mathbf{B}_{t_0}^{t}(\mathbf{p})\vert_{\eta=0} = \mathbf{0}.
\end{equation}
Using \eqref{eq:weingartenComplete}, \eqref{eq:InvariantFlatWall}, \eqref{eq:Bzero}, and assuming a flat no-slip wall, we obtain
\begin{equation}\label{eq:CompLagrSpike}
    \partial_{\eta}\overline{\mathbf{W}}_{t_0}^{t}(\mathbf{p})\vert_{\eta=0} = \left (\begin{smallmatrix}
    \int^{t}_{t_0}\partial_{uu\eta}f_3(\mathbf{r}_{\eta}(\mathbf{p}),\tau)\vert_{\eta=0}d\tau&\int^{t}_{t_0}\partial_{vu\eta}f_3(\mathbf{r}_{\eta}(\mathbf{p}),\tau)\vert_{\eta=0}d\tau\\
		\int^{t}_{t_0}\partial_{uv\eta}f_3(\mathbf{r}_{\eta}(\mathbf{p}),\tau)\vert_{\eta=0}d\tau&\int^{t}_{t_0}\partial_{vv\eta}f_3(\mathbf{r}_{\eta}(\mathbf{p}),\tau)\vert_{\eta=0}d\tau
    \end{smallmatrix}\right ).
\end{equation}
A similar expression can be obtained for curved boundaries.
\subsection{Incompressible flows}\label{App:LagrSpikingPointIncompr}
In the case of incompressible flows, by differentiating the continuity equation and using the no-slip condition on the wall, we obtain
\begin{equation}
    \partial_{uu}(\partial_u f_1+\partial_v f_2 + \partial_\eta f_3 )\vert_{\eta=0} = 0 \rightarrow \partial_{uu\eta}f_3\vert_{\eta=0} = 0. 
\end{equation}
Similarly, we can get $\partial_{uv\eta}f_3\vert_{\eta=0} = \partial_{vv\eta}f_3\vert_{\eta=0}=0 $, therefore we have,
\begin{equation}
    \partial_{\eta}\overline{\mathbf{W}}_{t_0}^{\tau}(\mathbf{p})\vert_{\eta=0} = \mathbf{0}.
\end{equation}
Therefore the leading order contribution in \eqref{eq:deltaEtaLagr} is $O(\delta\eta^2)$, which is given by 
\begin{equation}
    \overline{\mathbf{W}}_{t_0}^{t}(\mathbf{r_{\delta\eta}}(\mathbf{p})) = \underbrace{\partial_{\eta\eta}\overline{\mathbf{W}}_{t_0}^{t}(\mathbf{r_{\eta}}(\mathbf{p}))\vert_{\eta=0}}_{\partial_{\eta\eta}\overline{\mathbf{W}}_{\eta}(\mathbf{p})\vert_{\eta=0}}\frac{\delta\eta^2}{2} + O(\delta\eta^3).
\end{equation}
We can use $\partial_{\eta\eta}\overline{\mathbf{W}}_{\eta}(\mathbf{p})\vert_{\eta=0}$ to compute the eigenvalues and eigenvectors of $\overline{\mathbf{W}}_{t_0}^{t}(\mathbf{r_{\delta\eta}}(\mathbf{p}))$ to the leading order in $\eta$. We express $\partial_{\eta\eta}\overline{\mathbf{W}}_{\eta}(\mathbf{p})\vert_{\eta=0}$ in terms of the spatial derivatives of Eulerian quantities as
\begin{equation}
    \partial_{\eta\eta}\overline{\mathbf{W}}_{t_0}^{t}(\mathbf{r_{\eta}}(\mathbf{p}))\vert_{\eta=0} = \partial_{\eta\eta}\int_{t_0}^{t} \dot{\overline{\mathbf{W}}}_{t_0}^{\tau}(\mathbf{p})\vert_{\eta=0}d\tau = \int_{t_0}^{t} \partial_{\eta\eta}\dot{\mathbf{W}}_{t_0}^{\tau}(\mathbf{p})\vert_{\eta=0}d\tau.
\end{equation}
Using the same arguments in Appendix \ref{App:LagrSpikingPoint}, we obtain
\begin{equation}\label{eq:InCompLagrSpike}
    \partial_{\eta\eta}\overline{\mathbf{W}}_{t_0}^{t}(\mathbf{p})\vert_{\eta=0} = \left (\begin{smallmatrix}
    \int^{t}_{t_0}\partial_{uu\eta\eta}f_3(\mathbf{r}_{\eta}(\mathbf{p}),\tau)\vert_{\eta=0}d\tau&\int^{t}_{t_0}\partial_{vu\eta\eta}f_3(\mathbf{r}_{\eta}(\mathbf{p}),\tau)\vert_{\eta=0}d\tau\\
		\int^{t}_{t_0}\partial_{uv\eta\eta}f_3(\mathbf{r}_{\eta}(\mathbf{p}),\tau)\vert_{\eta=0}d\tau&\int^{t}_{t_0}\partial_{vv\eta\eta}f_3(\mathbf{r}_{\eta}(\mathbf{p}),\tau)\vert_{\eta=0}d\tau
    \end{smallmatrix}\right ).
\end{equation}

\section{Eulerian spiking point and curves}\label{app:SepPtOnWallEul}
We derive analytical expressions for the Eulerian spiking point and curves similar to their Lagrangian counterparts in Appendix \ref{App:LagrSpikingPoint}. The Eulerian spiking point and curve is defined as the point or curve that connects the Eulerian backbone of separation $\mathcal{B}(t)$ with the wall. Because the wall is an invariant set, we obtain 
\begin{equation}
    \dot{\mathbf{W}}_{t_0}^{t}(\mathbf{r_{\delta\eta}}(\mathbf{p})) = \underbrace{\partial_{\eta}\dot{\mathbf{W}}_{t_0}^{t}(\mathbf{r_{\eta}}(\mathbf{p}))\vert_{\eta=0}}_{\partial_{\eta}\dot{\mathbf{W}}_{\eta}(\mathbf{p})\vert_{\eta=0}}\delta\eta + O(\delta\eta^2).
\end{equation}
To determine the Eulerian spiking point and curves, similar to Appendix \ref{App:LagrSpikingPoint}, we derive analytical expressions for the leading order terms in $\eta$ of $\dot{\mathbf{W}}_{t_0}^{t}(\mathbf{r_{\eta}}(\mathbf{p}))$ at $\vert_{\eta=0}$ for compressible and incompressible flows as 
\begin{equation}\label{eq:CompEulSpike}
   \text{compressible:\quad} \partial_{\eta}\dot{\mathbf{W}}_{t_0}^{t}(\mathbf{p})\vert_{\eta=0} = \left (\begin{smallmatrix}
 \partial_{uu\eta}f_3(\mathbf{r}_{\eta}(\mathbf{p}),t)\vert_{\eta=0}&\partial_{vu\eta}f_3(\mathbf{r}_{\eta}(\mathbf{p}),t)\vert_{\eta=0}\\
		\partial_{uv\eta}f_3(\mathbf{r}_{\eta}(\mathbf{p}),t)\vert_{\eta=0}&\partial_{vv\eta}f_3(\mathbf{r}_{\eta}(\mathbf{p}),t)\vert_{\eta=0}
    \end{smallmatrix}\right ),
\end{equation}
\begin{equation}\label{eq:InCompEulSpike}
   \text{incompressible:\quad}  \partial_{\eta\eta}\dot{\mathbf{W}}_{t_0}^{t}(\mathbf{p})\vert_{\eta=0} = \left (\begin{smallmatrix}
    \partial_{uu\eta\eta}f_3(\mathbf{r}_{\eta}(\mathbf{p}),t)\vert_{\eta=0}&\partial_{vu\eta\eta}f_3(\mathbf{r}_{\eta}(\mathbf{p}),t)\vert_{\eta=0}\\
		\partial_{uv\eta\eta}f_3(\mathbf{r}_{\eta}(\mathbf{p}),t)\vert_{\eta=0}&\partial_{vv\eta\eta}f_3(\mathbf{r}_{\eta}(\mathbf{p}),t)\vert_{\eta=0}
    \end{smallmatrix}\right ).
\end{equation}
Comparing \eqref{eq:InCompEulSpike} and \eqref{eq:CompEulSpike} with \eqref{eq:InCompLagrSpike} and \eqref{eq:CompLagrSpike}, shows that for steady flows, the Eulerian spiking point and curve coincide with the Lagrangian spiking point and curve. 
\section{Creeping flow around a rotating cylinder}\label{App:Creepingflow}
\citet{klonowska2001exact} derived an analytical solution of a creeping flow around a fixed rotating circular cylinder close to an infinite plane wall moving at a constant velocity (Fig. \ref{fig:CreepingFlowSetup}). If $u$ and $v$ denote the velocity components along and normal to the wall, the solution is given by the following complex function 
\begin{equation}
\begin{aligned}
u(\zeta)-iv(\zeta)=&-\frac{U_w}{2\log a}\bigg{[}2\log\frac{\vert\varphi\vert}{a}+\frac{\mu}{2\varphi}(\zeta^*-\zeta)(\varphi-1)^2\bigg{]}\\
&+\sigma(\varphi-1)^2\bigg{[}\frac{i\mu\zeta^*}{2}\bigg{(}\frac{a}{\varphi^2}+\frac{1}{a}\bigg{)}-\frac{1}{\varphi}\bigg{(}a+\frac{1}{a}\bigg{)}+\frac{1}{2a}\bigg{(}\frac{a^2}{\varphi^2}-1\bigg{)}\bigg{]}\\ 
&+\sigma\bigg{[}a+\frac{1}{a}+i\bigg{(}\frac{a}{\varphi^*}-\frac{\varphi^*}{a}\bigg{)}\bigg{]},\\ 
\end{aligned}
\label{eq:complexFunctCreepingFlow}
\end{equation}
where 
\begin{equation}
i=\sqrt{-1},\ \ \zeta=x+iy,\ \ \varphi=\varphi(\zeta)=\frac{1+i\mu\zeta}{1+\mu\zeta},
\label{eq:complexFunctCreepingFlowIngred1}
\end{equation}
with $(\cdot)^*$ denoting the complex conjugate operator. The constants $a,\mu$ and $\sigma$ describe the geometry and the kinematics of the cylinder, and are defined as 
\begin{equation}
a=\frac{R_c+y_c-\sqrt{y_c^2-R_c^2}}{R_c+y_c+\sqrt{y_c^2-R_c^2}},\ \ \mu=\frac{1}{\sqrt{y_c^2-R_c^2}},\ \ \sigma=\frac{a}{a^2-1}\bigg{(}-\frac{U_w}{2\log a}+\frac{2\Omega a^2}{\mu(a^2-1)^2}\bigg{)}.
\label{eq:complexFunctCreepingFlowIngred2}
\end{equation}
In eq. \eqref{eq:complexFunctCreepingFlowIngred2} $U_w$ denotes the velocity of the wall and $R_c$ the radius of the cylinder initially centered at $(0,y_c)$, and rotating about its axis with angular velocity $\Omega$. Following the procedure described in \citet{miron2015towards}, by the linearity of Stokes flows, substituting $x$ and $u$ in eq. \eqref{eq:complexFunctCreepingFlow} with $x-U_wt-\tfrac{\beta}{\omega_c}\cos(\omega_ct)$ and $u-U_w$, we obtain the velocity field developing close to a rotating cylinder, whose centers moves parallel to a fixed wall with velocity $U_c=U_0+\beta\cos(\omega t)$, where $U_0=-U_w$.
	\section{Steady flow past a mounted cube}\label{sec:SteadyFlowCubeDetails}
    We consider the flow around a cube placed on a wall located at $x_3=0$, with $h=1$ the length of the edge of the cube. The finite difference code \texttt{Xcompact3d} \citep{Laizet2009,Laizet2011} was used to solve the incompressible Navier--Stokes equations at a Reynolds number $Re = U_{\infty}h/\nu = 200$, based on the free-stream velocity $U_{\infty}$, $h$ and the kinematic viscosity $\nu$. The computational domain is $L_{x_1}  \times L_{x_2} \times L_{x_3} = 12h \times 3h \times 3h$, where $x_1$, $x_2$ and $x_3$ are the streamwise, spanwise and transverse direction, respectively. The cube is centered on $x_1=4.5h$ and $x_2=1.5h$. The domain is discretized on a Cartesian grid (stretched in the $x_3$-direction) of $481\times 129 \times 129 $ points, with a 6th order finite difference compact scheme in space, while the time integration is performed with a 3rd order Adams-Bashforth scheme with a time step $\Delta t = 5 \times 10^{-5}h/U_{\infty}$.  A specific immersed boundary method is used to model the solid cube and to impose a no-slip condition on its faces \citep{Gautier2014}. At the bottom wall, a conventional no-slip condition is imposed, while at the top and lateral walls, a free-slip condition is chosen. A laminar Blasius velocity profile is prescribed at the inlet section, with a boundary layer thickness of $h/4$. Finally, at the outlet, a convective equation is solved.
    
    \section{Laminar separation bubble flow}\label{App:LSB}
	\textcolor{black}{The flat-plate LSB flow is computed using a high-order discontinuous Galerkin spectral element method \citep{kopriva,Klose2020} for the spatial discretization of the compressible Navier-Stokes equations and explicitly advanced in time with a 4th-order Runge-Kutta scheme. 
	The solution is approximated on a 7th order Legndre-Gauss polynomial basis yielding an 8th order accurate scheme.
	The computational domain is $L_{x_1}  \times L_{x_2} \times L_{x_3} = 200\delta_{\mathrm{in}}^* \times 10\delta_{\mathrm{in}}^* \times 15\delta_{\mathrm{in}}^*$, discretized with a total of 9,600 high-order elements (total of 4,915,200 degrees of freedom per equation).
	The Reynolds number is $Re_{\delta_{\mathrm{in}}^*} = U_\infty \delta_{\mathrm{in}}^*/\nu = 500$, based on the free-stream velocity $U_\infty$, the height of the inflow boundary layer displacement thickness $\delta_{\mathrm{in}}^*$ and the kinematic viscosity $\nu$. 
	The free-stream Mach number is 0.3. 
	At the inlet and outlet, a Blasius profile, corrected for compressible flow, is prescribed and the wall is set to be isothermal.
	The inflow profile is modified by $U_{in} = U_{Blasius}(1 + 0.1\cos{(2\pi x_2/L_{x_2})})$ to induce a spanwise modulation of the LSB.
	To avoid spurious oscillations at the outflow boundary, a spectral filter is applied for elements $x_1/\delta_{\mathrm{in}}^*>160$.
	A modified free-stream condition is prescribed at the top boundary, with a suction profile for the lateral velocity component $S(x_1)/U_\infty=a_s\exp{[-b_s(x_1/\delta_{\mathrm{in}}^*-c_s)]}$, according to \citet{AlamSandham2000}, and a zero-gradient for the streamwise component.
	The coefficients for the steady case are $a_s=1/5$, $b_s=1/50$ and $c_S=25$. 
	For the unsteady case, the constant coefficient $c_s$ is replaced by a time-dependent function $c_s(t)=30 + 5 \sin{(2\pi f t)}$ with $f=1/20$ to induce a periodically oscillating LSB.}
\end{appendices}
\bibliographystyle{ieeetr}
\bibliography{output.bbl}
\end{document}